\documentclass[aps,fleqn, pra, superscriptaddress, floatfix, showpacs, 10pt]{revtex4-1}
\pdfoutput=1
\usepackage{dcolumn}
\usepackage{bm}
\usepackage{amssymb, amsmath}
\usepackage{graphicx, float}
\usepackage{epstopdf}
\usepackage{subfigure}
\usepackage{color}
\usepackage{braket}
\usepackage{ulem}
\def\be{\begin{equation}}
\def\ee{\end{equation}}

\begin{document}

\title{Vortex creation without stirring in coupled ring resonators with gain and loss}

\author{Aleksandr Ramaniuk}
\affiliation{Faculty of Physics, University of Warsaw, ul. Pasteura 5,
PL-02-093 Warszawa, Poland}
\author{Nguyen Viet Hung}
\affiliation{Advanced Institute for Science and Technology, Hanoi
University of Science and Technology, 100803 Hanoi, Vietnam}
\author{Michael Giersig}
\affiliation{Department of Physics, Freie Universit\"at Berlin, Arnimalle 14,
D-14195 Berlin, Germany}
\affiliation{International Academy of Optoelectronics at Zhaoqing, South China Normal University, 526238 Guangdong, P. R. China}
\author{Krzysztof Kempa}
\affiliation{Boston College, Department of Physics, Chestnut Hill, MA
02467 USA}
\author{Vladimir V. Konotop}
\affiliation{Centro de F\'isica Te\'orica
e Computacional and Departamento de
F\'isica, Faculdade de Ci\^encias, Universidade de Lisboa, Campo
Grande, Edif\'icio C8, Lisboa 1749-016,  Portugal}
\author{Marek Trippenbach}
\affiliation{Faculty of Physics, University of Warsaw, ul. Pasteura 5,
PL-02-093 Warszawa, Poland}
\date{\today}


\begin{abstract}
We present study of the dynamics of two ring waveguide structure with space dependent coupling, { linear} gain and { nonlinear} absorption - the system
that can be implemented in polariton condensates, optical waveguides, and nanocavities. We show that by turning on and off local coupling between rings one can selectively generate permanent vortex in one of the rings. We find that due to the modulation instability it is also possible to observe several complex nonlinear phenomena, including spontaneous symmetry breaking, stable inhomogeneous states with interesting structure of currents flowing between rings, generation of stable symmetric and asymmetric circular flows with various vorticities, etc. The latter can be created in pairs (for relatively narrow coupling length) or as single vortex in one of the channels, that is later alternating between channels.
\end{abstract}

\maketitle
\setcounter{section}{0}

\section{Introduction}

Coupled microrings (microdiscs, or more generally microcavities) are
standard basic elements in diverse physical applications.  In optics
they are used for nonreciprocal devices~\cite{nonrecip_optics},
switches~\cite{switches}, loss control of lasing~\cite{lasing}, and
ring lasers~\cite{laser1,laser2},  to mention a few. Recently the
attention was also turned to more sophisticated devices which
include several coupled microcavities and which are referred to as
photonic molecules~\cite{phot_molec}. Coupled non-Hermitian
microcavities are also used for study of chiral modes in
exciton-polariton condensates~\cite{ex-polar-BEC}, as well as  for
modeling coupled circular traps for Bose-Einstein condensates (BEC), where
gain corresponds to adding atoms while nonlinear losses occurs due
to inelastic two-body interactions. They can also be realized in
nanoplasmonic systems \cite{wang} and slow-light optical microcavities \cite{huet}.

Like in case of any coupled subsystems, the characteristics of
coupling between  microrings (or to an external element, for
instance to a bus fiber) may strongly affect the stationary regimes
as well as the dynamics supported by the system. The coupling can be
modified in various ways. It depends on the geometry (i.e. on the
mutual locations of the rings), on the wave-guiding characteristics
of the rings which determine the field decay outside the cavities,
on the medium between the cavities (it can be homogeneous or
gradient; active, absorbing, or conservative), etc. Thus, it is of
natural interest to understand how the characteristics of coupling
affect the field distribution and dynamics inside the ring cavities.
This is the question which is addressed in the present work. The
main emphasis is made on the interplay between the size of the
coupling domain and other spatial scales of the system (i.e. on the
ring lengths and on the characteristic scales of the excited modes).

Before we move on to specific model (formulated in
Sec.~\ref{sec:model}) whose stationary solutions are investigated in
Sec.~\ref{sec:stat}, the dynamical regimes in Sec.~\ref{sec:narrow}
and  Sec.~\ref{sec:broad}, and vortices in~Sec.~\ref{sec:vortex}, we
would like to advertise our effort to design up a nanostructure of
coupled rings to explore experimentally this and similar settings
and its dynamics. In Sec.~\ref{sec:experiment} we discuss specific
experimental proposal based on the structures that were already
manufactured.

\section{The model
}
\label{sec:model}

In the present study we consider a model described by two coupled
nonlinear Schr\"odinger equations with gain and nonlinear loss
(depending on applications they also can be termed Gross-Pitaevski
or Ginzburg-Landau equations), which we write down in scaled
dimensionless units
\begin{eqnarray}
\label{eq:1}
\begin{array}{l}
i \partial_t \psi_1 = -\partial_x^2 \psi_1 + i \gamma \psi_1 + (1 - i \Gamma) \vert \psi_1 \vert^2 \psi_1 + J(x) \psi_2 ,
\\[1mm]
i \partial_t \psi_2 = -\partial_x^2 \psi_2 + i \gamma \psi_2 + (1 - i \Gamma) \vert \psi_2 \vert^2 \psi_2 + J(x) \psi_1.
\end{array}
\end{eqnarray}
Here $\psi_1$ and $\psi_2$ are the fields in the first and second waveguides, $\gamma$ is the linear gain and $\Gamma$  is the nonlinear loss, both considered constants along the waveguides, and $J(x)$ is position depending coupling.

Extended discussion of  model (\ref{eq:1}) and of its applications can be found in a previous publication \cite{SciRep} where the rings were homogeneously coupled, i.e. where it was assumed that $J(x)$ is constant. We also mentioned that  (\ref{eq:1}) with constant coupling is analogous to the model introduced earlier in~\cite{Malomed} where it was considered on the whole axis subject to the zero boundary conditions. In this paper we focus on expanding the study of the model through introducing coupling modulation $J(x)$. We consider rings assuming, without loss of generality, that $x\in[-\pi,\pi]$. This implies periodic boundary conditions for both channels:  $\psi_i (x,t) = \psi_i (x+2\pi,t)$, and the coupling function $J(x)$ is extended only in the certain region of the rings. In particular, for numerical simulations we shall consider local Gaussian coupling in the following form
\begin{equation} \label{eq:2}
J(x) =  \dfrac{J_0}{\sqrt{\pi} w} \exp\left( -\dfrac{x^2}{w^2}
\right)
\end{equation}
where $w$ is the width of the coupling, while $J_0$ characterizes the coupling strength. Our results are not sensitive to the particular shape of the wavefunction, as we have checked using supergaussian functions with very high power.

  An important remark about the used terminology is in order. For all applications mentioned in the Introduction, the meaning of the variable $x$ is an angle defining a point on the circumference. The functions $\psi_{1,2}$ are rather  {\em  envelopes} of the field distributions than the total fields (see e.g.~\cite{Menyuk} for optical resonators and \cite{SaitoUeda,BluKon} for BEC applications). Thus, solutions for $\psi_{1,2}$ having nonzero topological charge (see \cite{SciRep}) may correspond to the total field distributions having phase singularities in the centers of the rings. In other words, such solutions describe {\em vortices}. Taking this into account the respective solutions are referred below as vortices.
System (\ref{eq:1}) is simple, but possesses surprisingly rich and diverse set of stable states (some of them nonstationary). For the limit of very wide coupling ($w \gg \pi$) we expect the same results dynamics as for the constant coupling (it is described in \cite{SciRep}). On the other side, very narrow coupling ($w \ll \pi$) allows to approximate coupling with delta function.

Most of the results found in the present study are numerical (using propagation techniques). In this context there is one  important issue that we need to address before we present the outcome of our investigations. As discussed in \cite{SciRep}, for uncoupled case ($J_0=0$) one can find stable background solutions in the form
\begin{equation} \label{eq:3}
\psi_{1,2}(t) = \sqrt{\dfrac{\gamma}{\Gamma}} e^{-i \dfrac{\gamma}{\Gamma} t}.
\end{equation}
Once the rings become coupled, modulation instability occurs mostly due to the interplay between gain and  nonlinear absorption.  In the case of constant coupling in Ref.~\cite{SciRep} two distinct classes of solutions have been found analytically: symmetric, characterized by $\psi_1=\psi_2$, and anti-symmetric with $\psi_1=-\psi_2$. The anti-symmetric solutions are always stable, and symmetric are usually unstable. Therefore we decided to perform numerical studies using {\it symmetric} state as initial condition. This led us to plethora of new states and attractors \cite{SciRep}.

\begin{figure}
    \includegraphics[scale=.33]{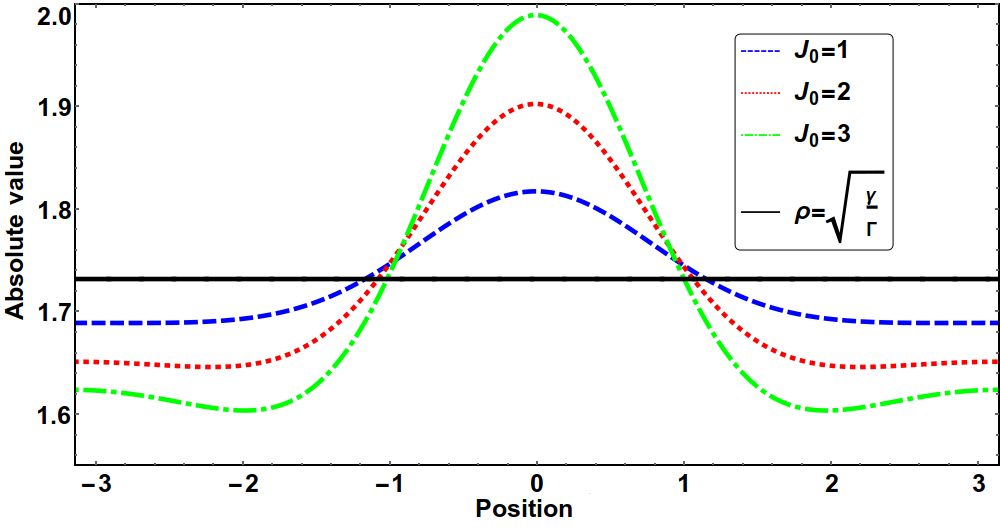}
    \includegraphics[scale=.33]{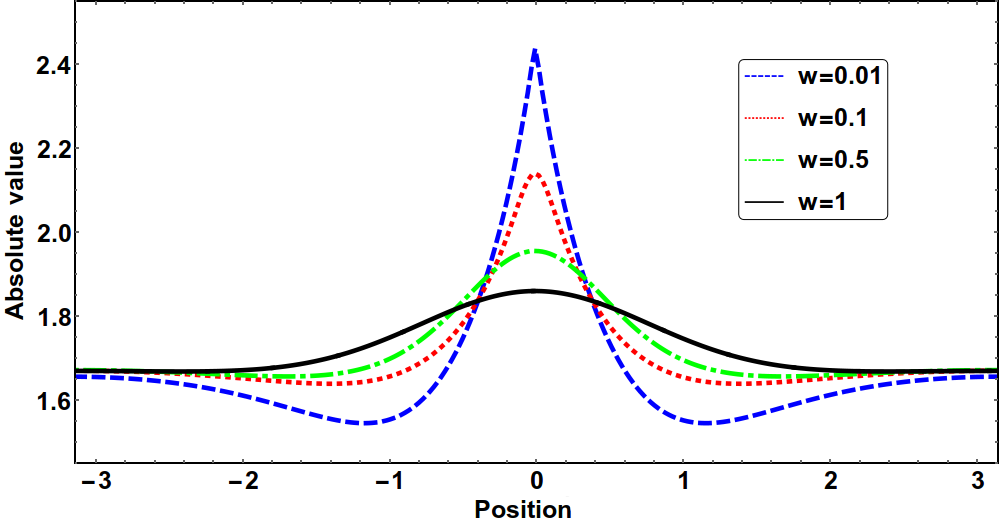}
    \caption{Absolute values of antisymmetric stationary states after propagation time $T=100$ in the coupled double-ring system (\ref{eq:1})  {obtained for the initial conditions (\ref{eq:4}) } with $\gamma=3$ and $\Gamma=1$. Left panel: Antisymmetric states calculated for different coupling strengths and fixed coupling width ($w=1$). Black line represents homogeneous state  for the respective uncoupled system. Right panel: Antisymmetric states calculated for different coupling widths and fixed normalized coupling ($J_0 = 1$).  }.
    \label{fig:1}
\end{figure}

We found that even if the coupling is not uniform, dynamics starting from anti-symmetric states lead to anti-symmetric {\it stationary} stable solution (see the examples of such attractors in Fig.~\ref{fig:1}). On the other hand, starting from a symmetric state, in some regions of parameters, does not necessarily lead to the antisymmetric stationary states, but can traverse to the more interesting attractors, like limit cycles or even chaotic states. Hence below we focus on the dynamics starting with the symmetric initial state with small perturbation imposed in the form of
\begin{equation} \label{eq:4}
\psi_{1,2} (x,t=0)= \sqrt{\dfrac{\gamma}{\Gamma}}(1 + \beta \sin (k x)),
\end{equation}
where the perturbation $\beta$ was typically of order of $10^{-2}$. In our simulations we took the value of the gain parameter $\gamma$ larger than loss $\Gamma$ and we checked that all results are qualitatively the same, regardless of the particular values of this parameters. The results also do not depend on particular values of the amplitude of the perturbation $\beta$ or perturbation wavenumber $k$. In the case of stronger loss ($\Gamma > \gamma$) results seem to be different and they are not included here. In this article we assume $\gamma=3, \Gamma=1$ for all later considerations and propagate from initial state defined with (\ref{eq:4}), unless stated otherwise. All simulations were performed via Split-Step Fourier method \cite{SSFM}.

\section{Stationary solutions}
\label{sec:stat}

When the coupling between the rings is weak ($J_0 \leq 1$), starting
from the initial state (\ref{eq:4}) we observed that the propagation
leads to stable,  stationary anti-symmetric solutions. Resulting
wavefunction has form of constant background with the bulge in the
coupling region, which depends on coupling function $J_0$, as
presented in Fig.~\ref{fig:1}. In this figure we plot the modulus of
the wavefunction for unitary width ($w=1$) and various coupling
constant $J_0$ (Fig.~\ref{fig:1}~left panel). In the right panel of
Fig.~\ref{fig:1} we fix $J_0=1$ and change $w$, going towards narrow
distributions, to show what one can expect when the coupling has a
form of Dirac delta function. We also show background level,
plotting it as a black horizontal reference line, in left panel of
Fig.~\ref{fig:1}.

In order to interpret the results of Fig.~\ref{fig:1}, we first
notice that for antisymmetric solutions $\psi_2=-\psi_1$ the
coupling plays the role of the linear potential well, $-J(x)$, thus
leading to the equation
\begin{align} \label{eq:11}
i \partial_t \psi_1 = -\partial_x^2 \psi_1  - J(x) \psi_1 + i \gamma \psi_1 + (1 - i \Gamma) \vert \psi_1 \vert^2 \psi_1
\end{align}
Let us now rewrite this model in the hydrodynamic form, introducing
the amplitude distribution $\rho(x)$ as well as the phase gradient
$v(x)=\partial_x \arg[\psi_1(x)]$, through the relation
$\psi_1(x)=\rho(x)\exp\left(i\int v(x)dx\right)$. For the stationary
solution we obtain from (\ref{eq:11}):
\begin{eqnarray}
\label{eq:im}
\frac{2v(x)}{\rho(x)}\frac{d\rho(x)}{dx}+\Gamma\rho^2(x)+\frac{dv(x)}{dx}-\gamma=0
\\
\label{eq:re}
\frac{1}{\rho(x)}\frac{d^2\rho(x)}{dx^2}-\rho^2(x)-v^2(x)+J(x)+\mu=0
\end{eqnarray}
Due to the parity symmetry of the problem $\psi_1(x)=\psi_1(-x)$ and
taking into account the continuity of the solution (i.e. of the
functions $\rho(x)$ and $v(x)$, and of their derivatives) we have
the relations.
\begin{eqnarray}
\rho_x(0)=\rho_x(\pm\pi)=0
\quad \mbox{and}\quad
v(0)=v(\pm \pi)=0
\end{eqnarray}

Now we observe that the maximal density $\rho_{{\rm max}}=\rho(0)$,
is achieved at $x=0$, i.e. in the point of the minimum of the
potential energy landscape $-J(x)$. Thus, at $x=0$ we obtain from
(\ref{eq:im}) that $\rho_{{\rm max}}=(\gamma-v_x(0))/\Gamma$, and
thus $v_x(0)<0$ meaning that the energy flow is directed towards the
minimum of the effective potential (what is expectable):
\begin{eqnarray}
v(x)>0\quad\mbox{at $x\in(-\pi,0)$}
\quad \mbox{and}\quad
v(x)<0\quad\mbox{at $x\in(0,\pi)$}
\end{eqnarray}
Similarly one can analyze the point $x=\pm\pi$, located at the
opposite side of the ring diameter. Denoting $\rho(\pm\pi)=\rho'$ we
have  $\rho'=(\gamma-v_x(\pm\pi))/\Gamma$. Now $v_x(\pm\pi)\geq 0$
and thus $\rho'\leq \rho_0=\sqrt{\gamma/\Gamma}$. This is what we
observe in both panels of Fig.~\ref{fig:1}. In particular, in the
left panel of Fig.~\ref{fig:1}  we observe decrease of $\rho'$ with
increase of $J_0/w$.

An interesting feature in the density distributions shown in both
panels of Fig.~\ref{fig:1} is the appearance of non-monotonic
dependence of $\rho(x)$ in the intervals $x\in(-\pi,0)$ and
$x\in(0,\pi)$. In particular, the minimal density is achieved in two
symmetric points $x=\pm x_{m}$ of the ring, different from
$x=\pm\pi$. As it follows from (\ref{eq:im}), at these points the
absolute value of the velocity gradient $v_x(\pm x_m)$ is maximal.
Since the  solution separating monotonic and non-monotonic densities
(in the intervals  $(-\pi,0)$ and  $(0,\pi)$)  is characterized by
$\rho_{xx}(\pm \pi)=0$ (at this solution the curvature of $\rho(x)$
changes its sign), it is not difficult to make an estimate of the
parameters for that solution in the case of sufficiently narrow
coupling. Indeed in that case $J(\pi)$ and $v_x(\pi)$ become
negligibly small and one obtains from  (\ref{eq:im}) and
(\ref{eq:re}) $\mu\approx\rho^2(\pi)\approx\gamma/\Gamma$.

In following subsections we present results of our studies of the
dynamics of our system with increasing coupling strength. We shall
distinguish two separate regimes of very narrow  ($w\ll 1$) and
extended ($w\gtrsim 1$) couplings. As we shall see in these limiting
cases the dynamics can be of a quite different character.

\section{Narrow coupling dynamics}
\label{sec:narrow}

\begin{figure}
\centering
\includegraphics[scale=.33]{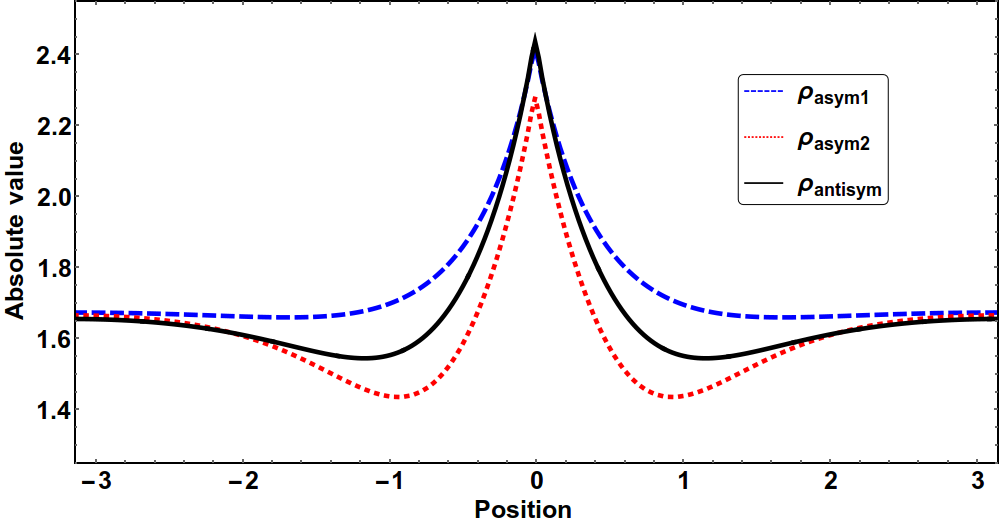}
\includegraphics[scale=.33]{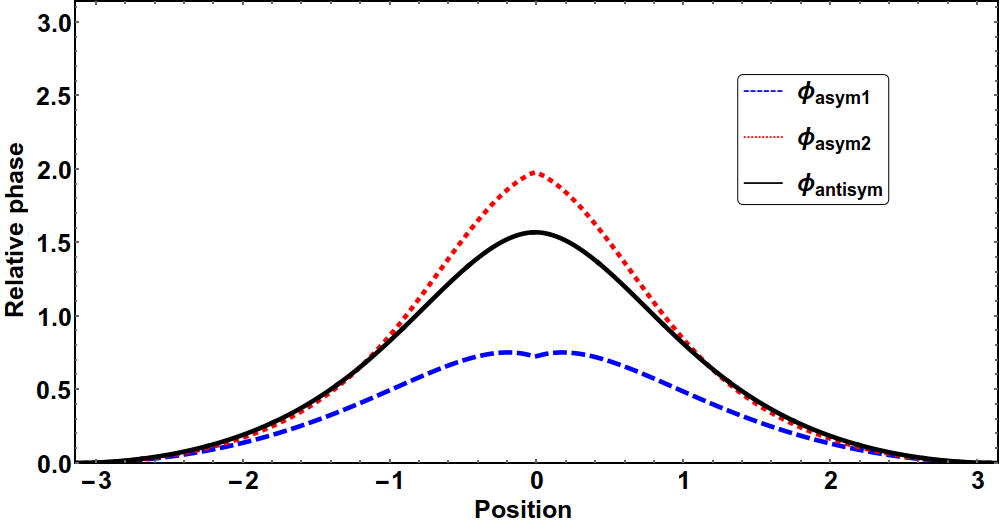}
\caption{Absolute value (left) and the phase (right) of stationary solutions observed in case of narrow coupling $w=0.01$. Black curve represents antisymmetric solution $ \psi_1(x)  = -\psi_2(x) $ ($J_0=1$), blue and red curves show absolute value of both channels in asymmetric state ($J_0=1.5$). Phase oscillating term is eliminated so that $\phi_1 (x=\pm \pi)=0$. Relative phase of the second channel in antisymmetric case is not shown, as $\phi_{2,antisym}=\phi_{1,antisym}+\pi$. Note that $\phi_{asym1}-\phi_{asym2}$ equals zero at $x=\pm\pi$.
} \label{fig:2}
\end{figure}

In this paragraph we present results obtained in simulations with very narrow coupling function $J(x)$, where we used Gaussian (\ref{eq:2}) of the width equal to $w=0.01$. Our results are summarized in Figs.~\ref{fig:2}, \ref{fig:3} and \ref{fig:4}. They are also described in the concise form below.

Depending on the value of $J_0$ we can distinguish several different patterns characteristic for the long (asymptotic) regular behavior. It can be stationary or periodic. In all simulations, the system reaches  respective attractor state after a transient period, which depends on initial conditions and perturbation, and typically doesn't exceed $t_{\rm trans}\approx 50$ in our arbitrary units.

For small coupling $(J_0 \lesssim 1)$ asymptotically long time dynamics lead to stable anti-symmetric state shown with the black curve in Fig.~\ref{fig:2}.  This state exhibits non-trivial phase structure, which is formed as a result of non-Hermiticity of our model.

For slightly larger coupling $(J_0 \simeq 1.5)$ when we propagate initial symmetric distribution for as long as it takes to reach steady state, we observe {\em  symmetry breaking} (between channels) and our system approaches a {\em stable   asymmetric} state, represented by blue ($\vert \psi_1 \vert$) and red ($\vert \psi_2 \vert$) curves in Fig.~\ref{fig:2}. We observe that local minima of the densities $\rho_{1,2}$ are achieved at $x=\pi$ for the first ring and for intermediate points $\pm x_m$ where $|x_m|<\pi$ for the second ring. To make a plot of the phase we excluded an $x-$independent component of the phase linearly growing in time. Note that hydrodynamic formulation allows us to discuss energy flow in the system. An interesting observation is that in the center of the coupling, i.e. in the vicinity $x=0$, the energy flows in the two rings have opposite directions. Indeed these flows are determined by $v_j$ and in the vicinity $x=0$ in the first ring with higher field density the current is directed outwards from the center ($x v_1>0$), while in the second ring it is directed towards the center ($x v_2<0$).

Finally, for larger coupling $(J_0 \gtrsim 2)$ our system tends to the limit cycle dynamics and we observe oscillations symmetrical with respect to the center of the coupling region. It is an interesting class of solutions, and analogous dynamics was found in the broad coupling regime (see the next paragraph). It seems to be universal and the general picture of the dynamics in this regime does not depend on the width of the coupling. Due to this universality we decided to discuss this class of solutions only for the broad coupling in the next paragraph.

\section{Broad coupling dynamics}
\label{sec:broad}

\begin{figure}
    \includegraphics[scale=.31]{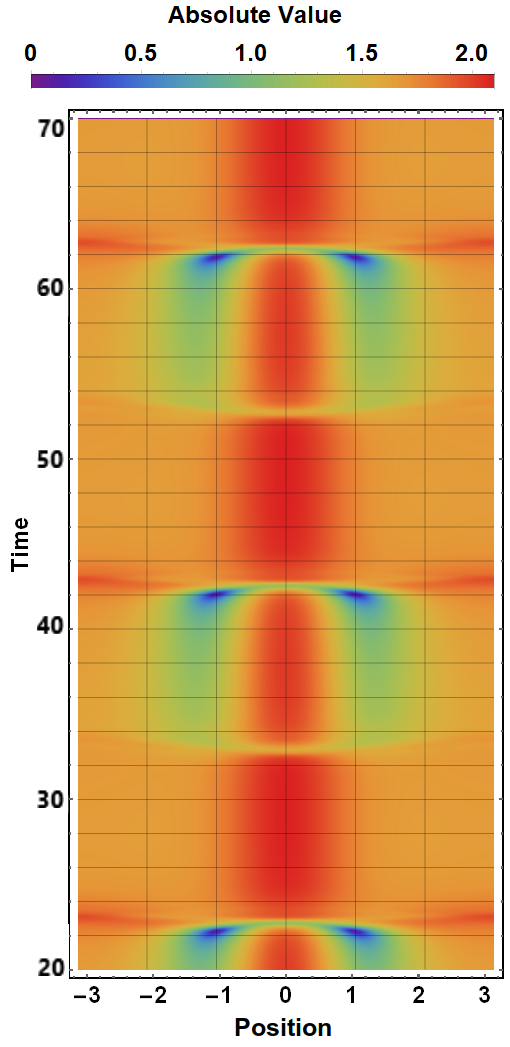}
    \includegraphics[scale=.31]{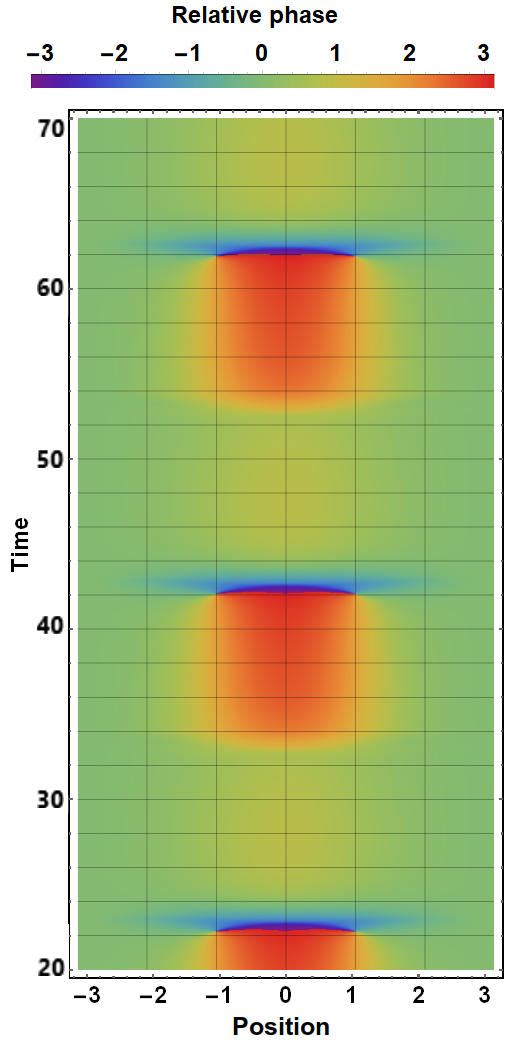}
    \includegraphics[scale=.31]{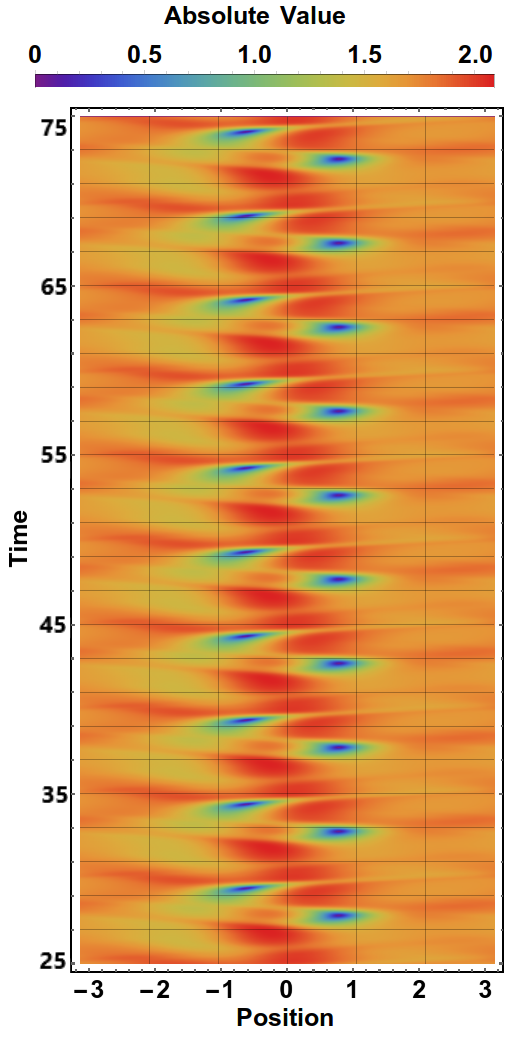}
    \includegraphics[scale=.31]{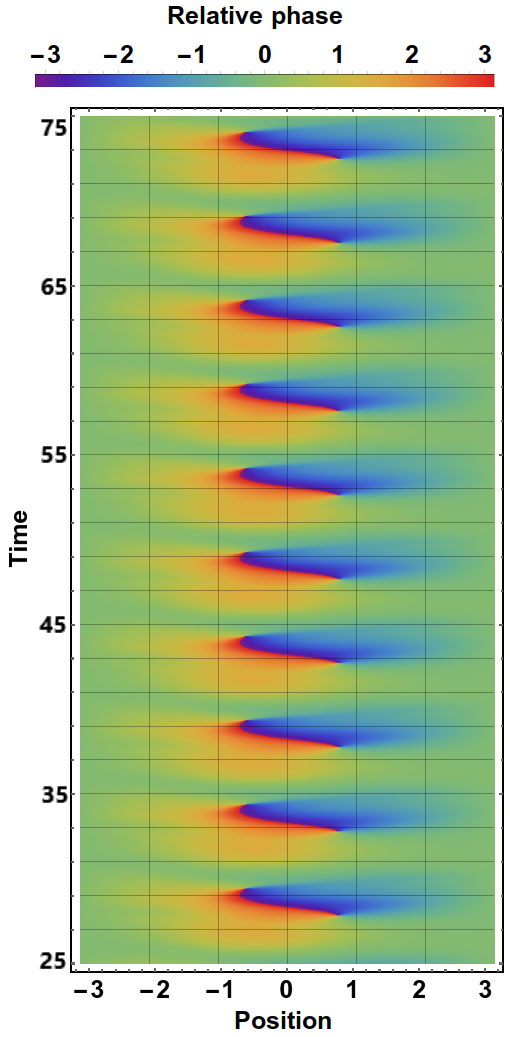} \\
    \includegraphics[scale=.325]{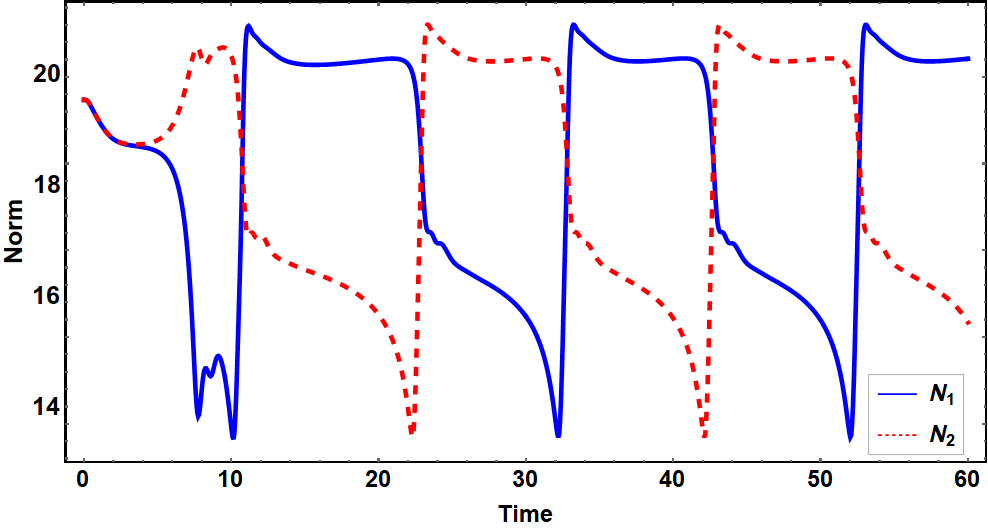}
    \includegraphics[scale=.325]{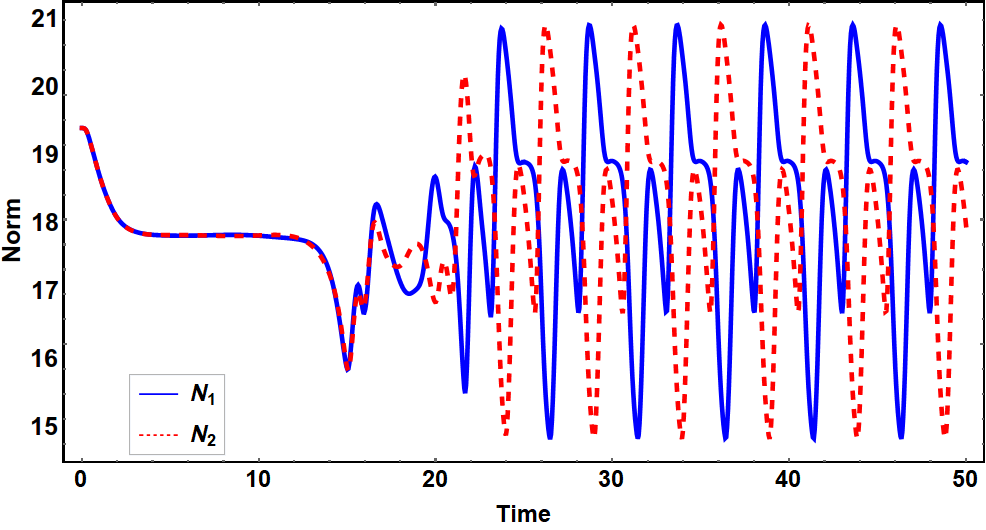} \\
    \includegraphics[scale=.243]{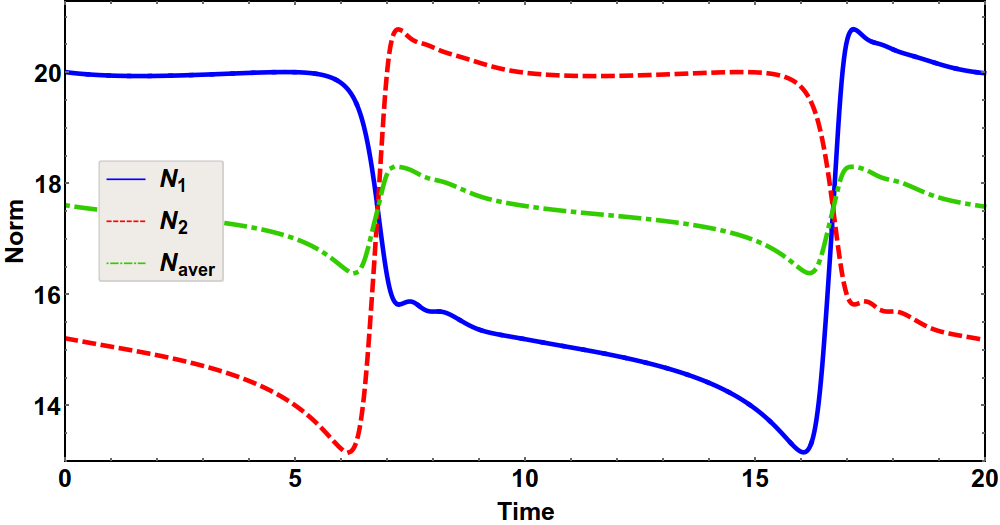}
    \includegraphics[scale=.243]{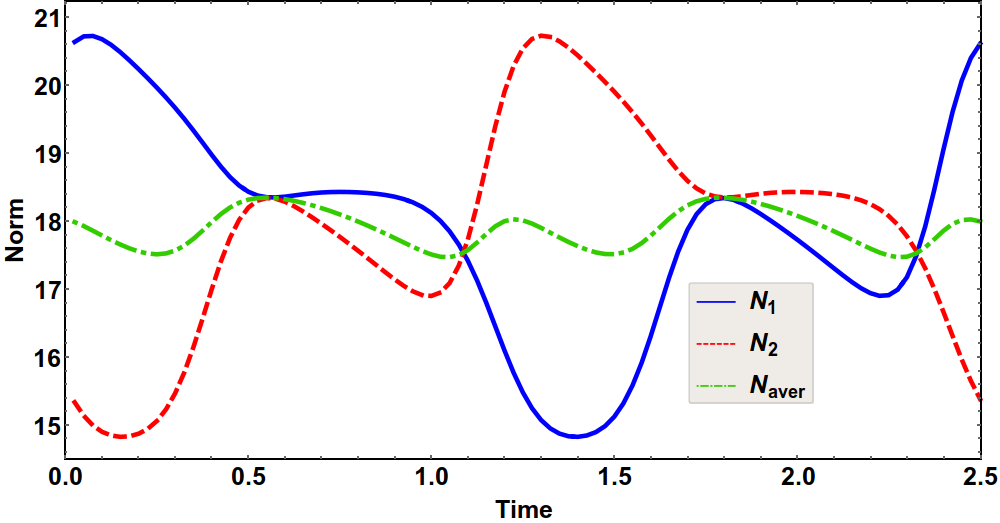}
    \caption{Top row: Contour plots of absolute values and phases of the propagated wavefunction $\psi_1$ in the stationary (asymptotic) regime for two different coupling stengths $J_0$ and the fixed width $w=1$. Phase oscillation term is eliminated so that $\phi_1 (x=\pi)=0$.
        Central row: Norms in both channels ($N_1$ and $N_2$) during propagation from initial perturbed symmetric homogeneous states, defined in Eq.~(\ref{eq:4}).
        Bottom row:  Norms of both channels and average $(N_1+N_2)/2$ during one period of oscillations in the limit cycle regime. Note that we shifted the time in order to show directly the length of the period of oscillations.
        Panels on the left show symmetric oscillations for $J_0=4$ and panels on the right show asymmetric oscillations for $J_0=5$.} \label{fig:3}
\end{figure}

\begin{figure}
    \centering
    \includegraphics[scale=.11]{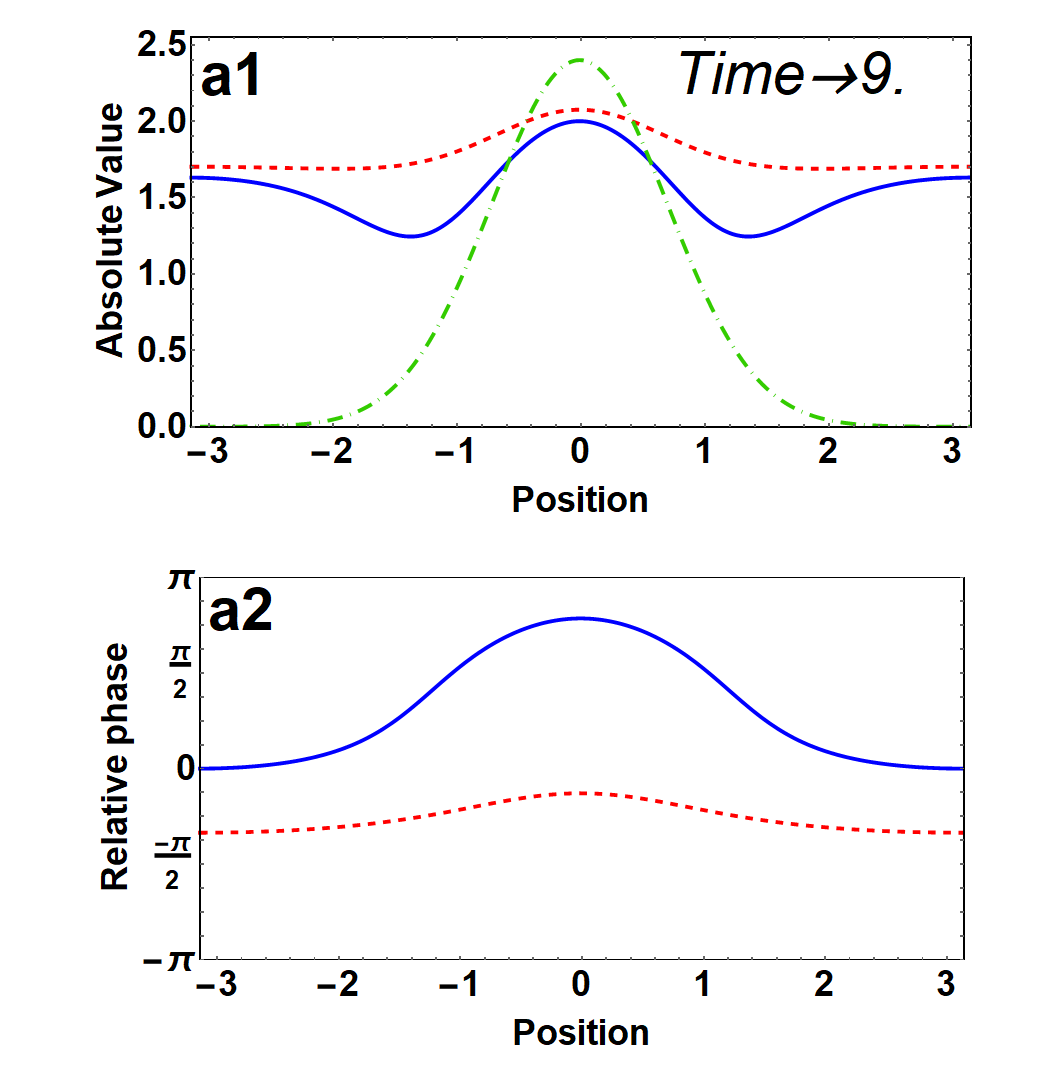}
    \includegraphics[scale=.11]{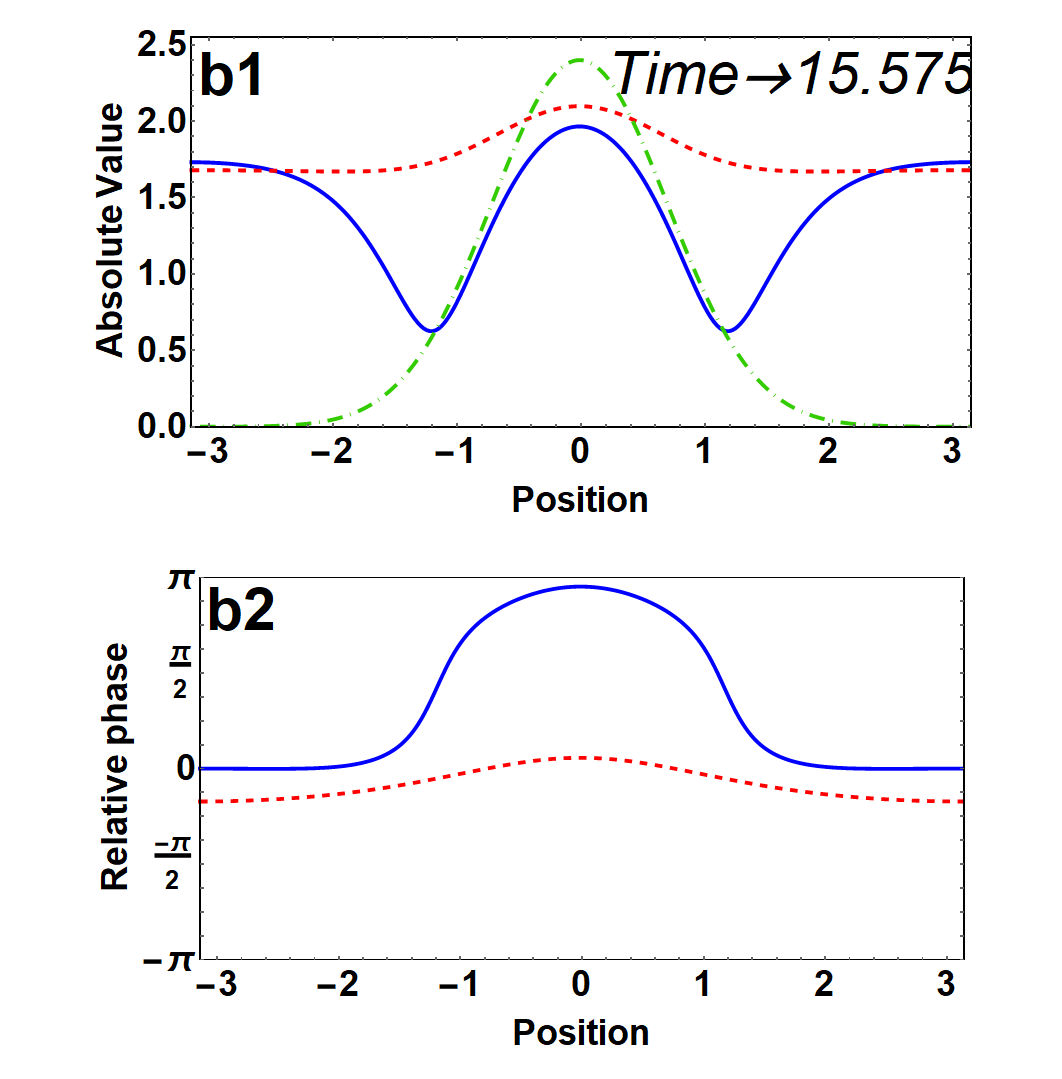}
    \includegraphics[scale=.11]{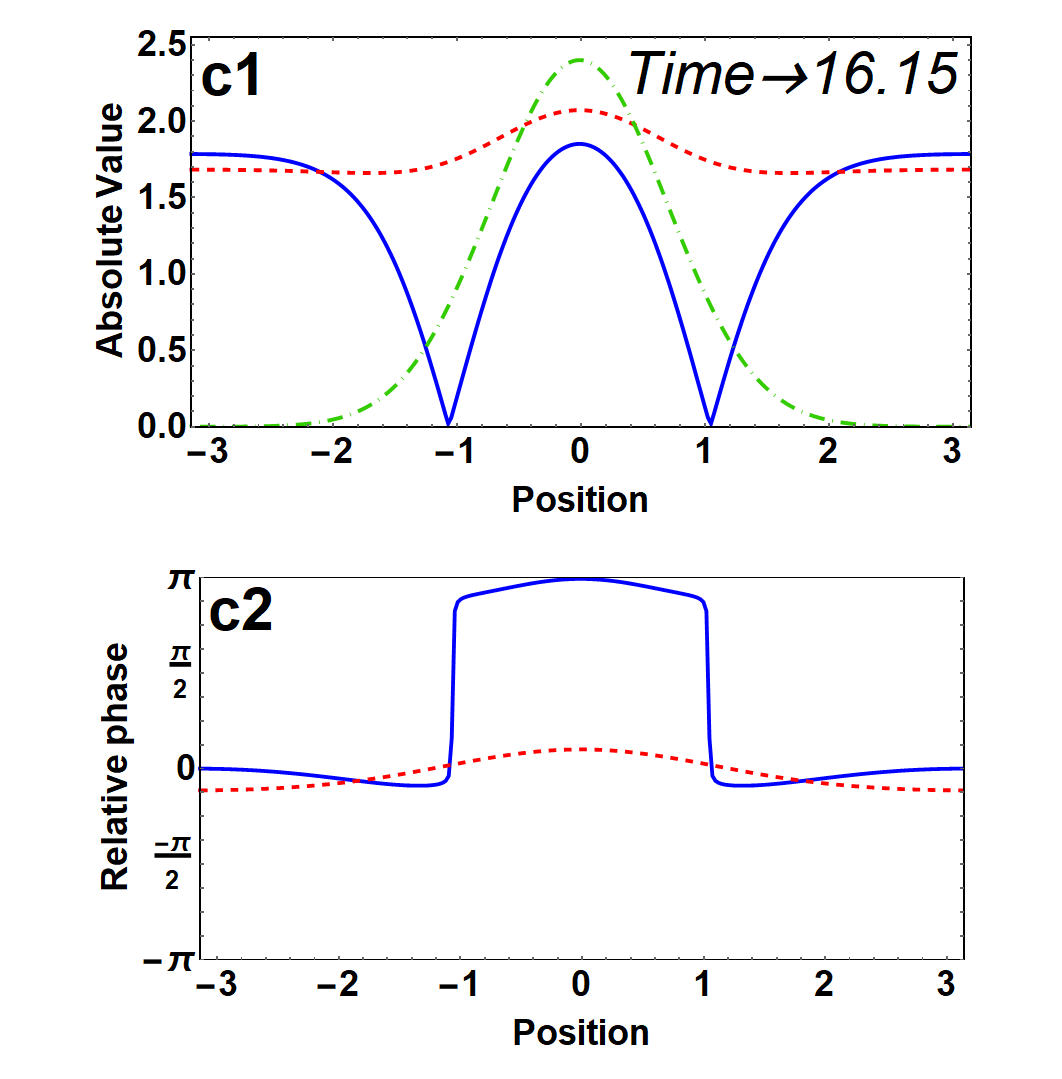}
    \includegraphics[scale=.11]{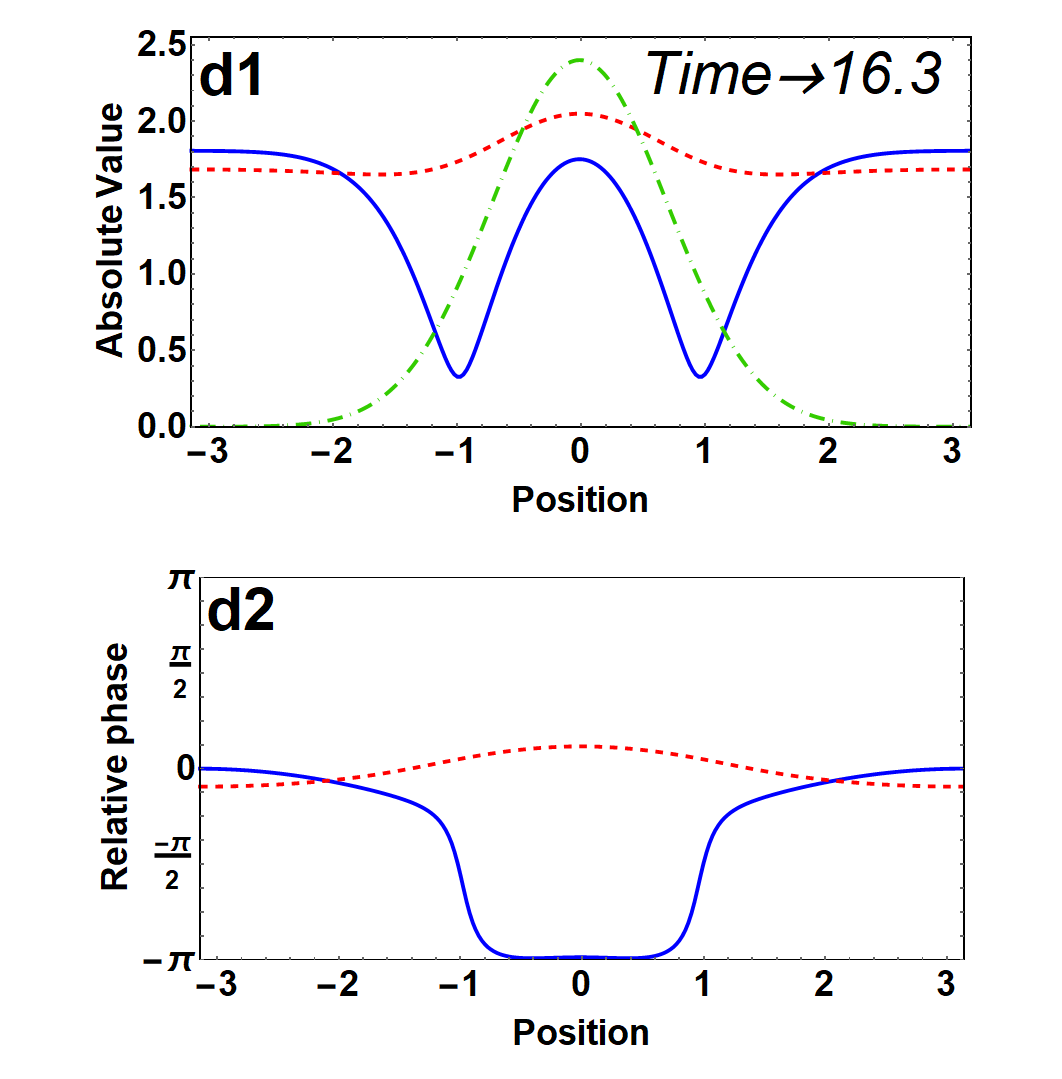}\\
    \includegraphics[scale=.11]{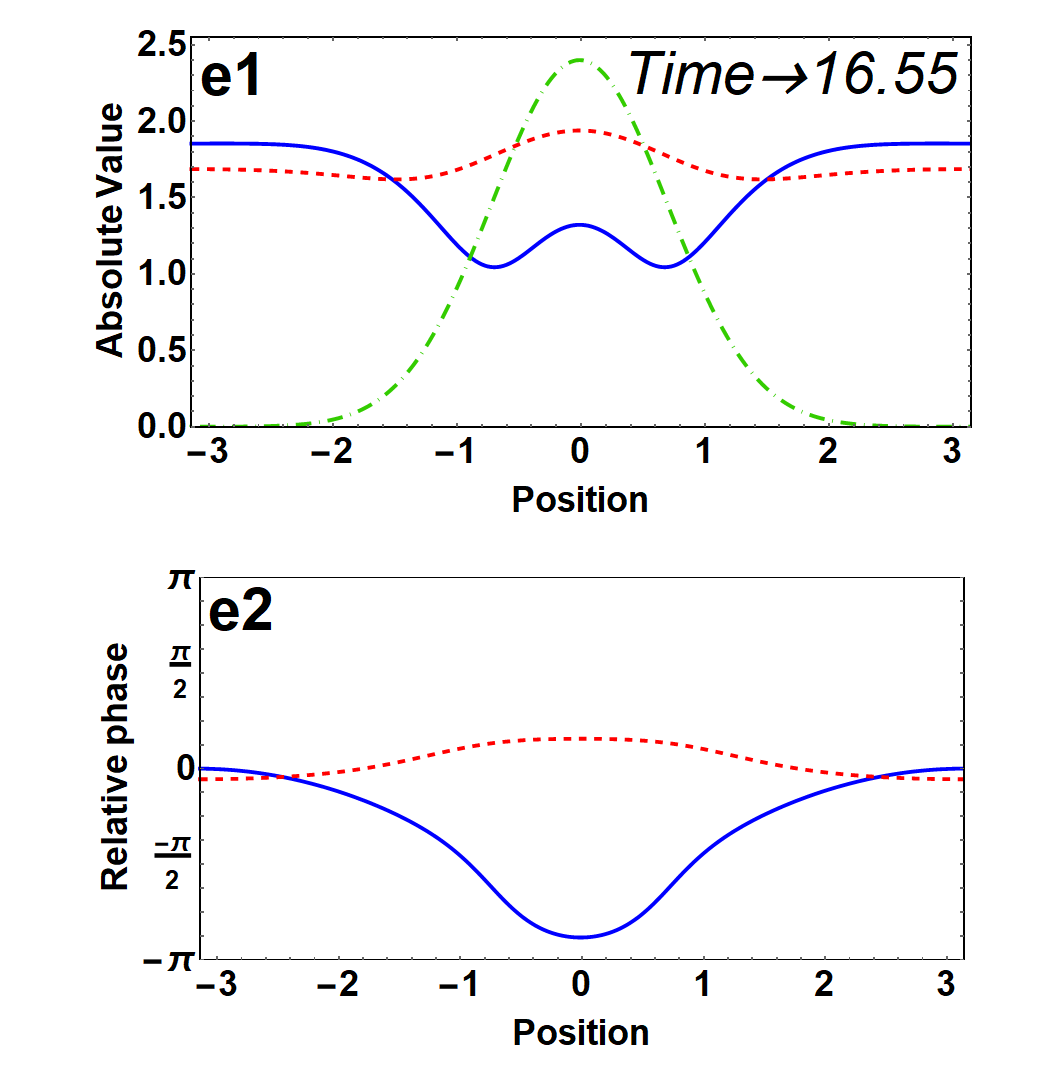}
    \includegraphics[scale=.11]{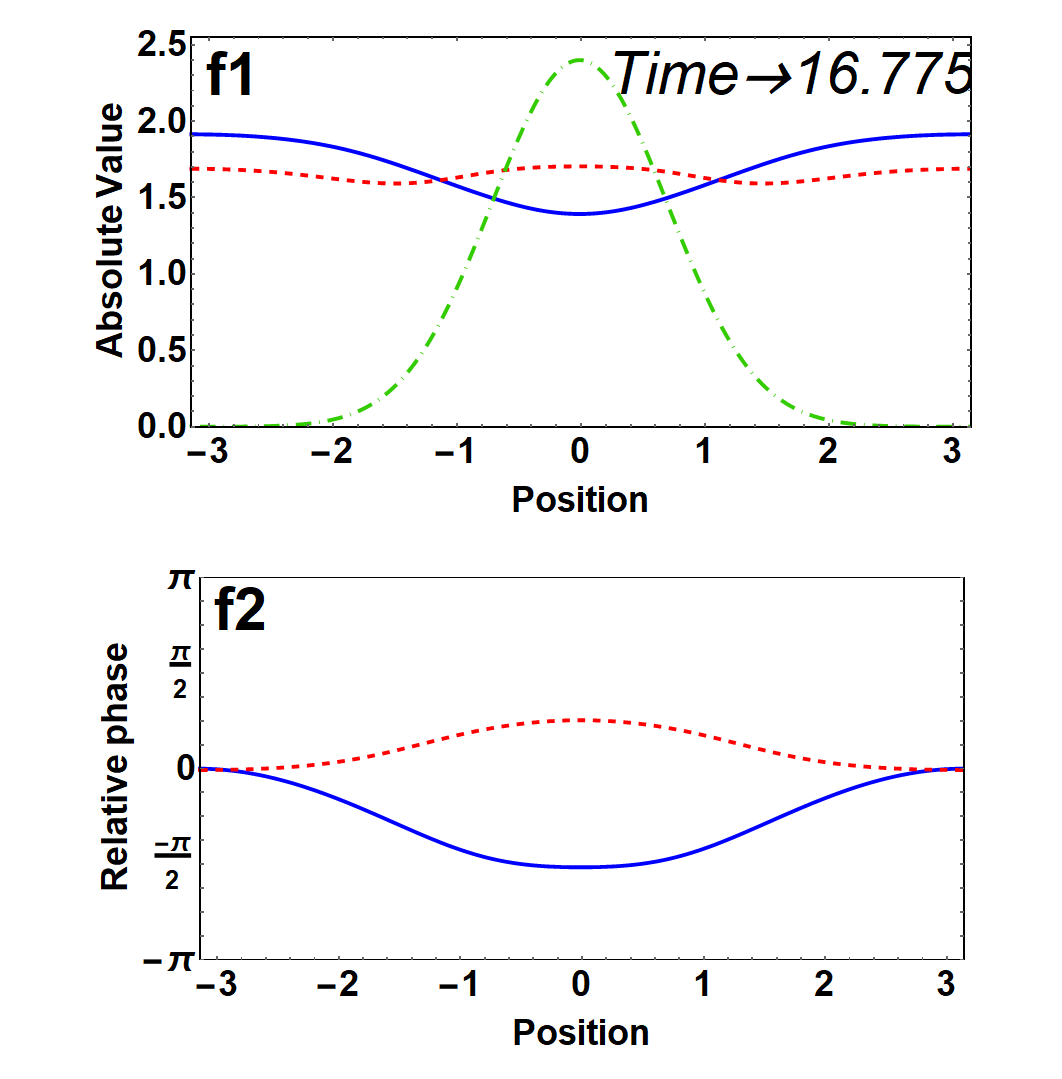}
    \includegraphics[scale=.11]{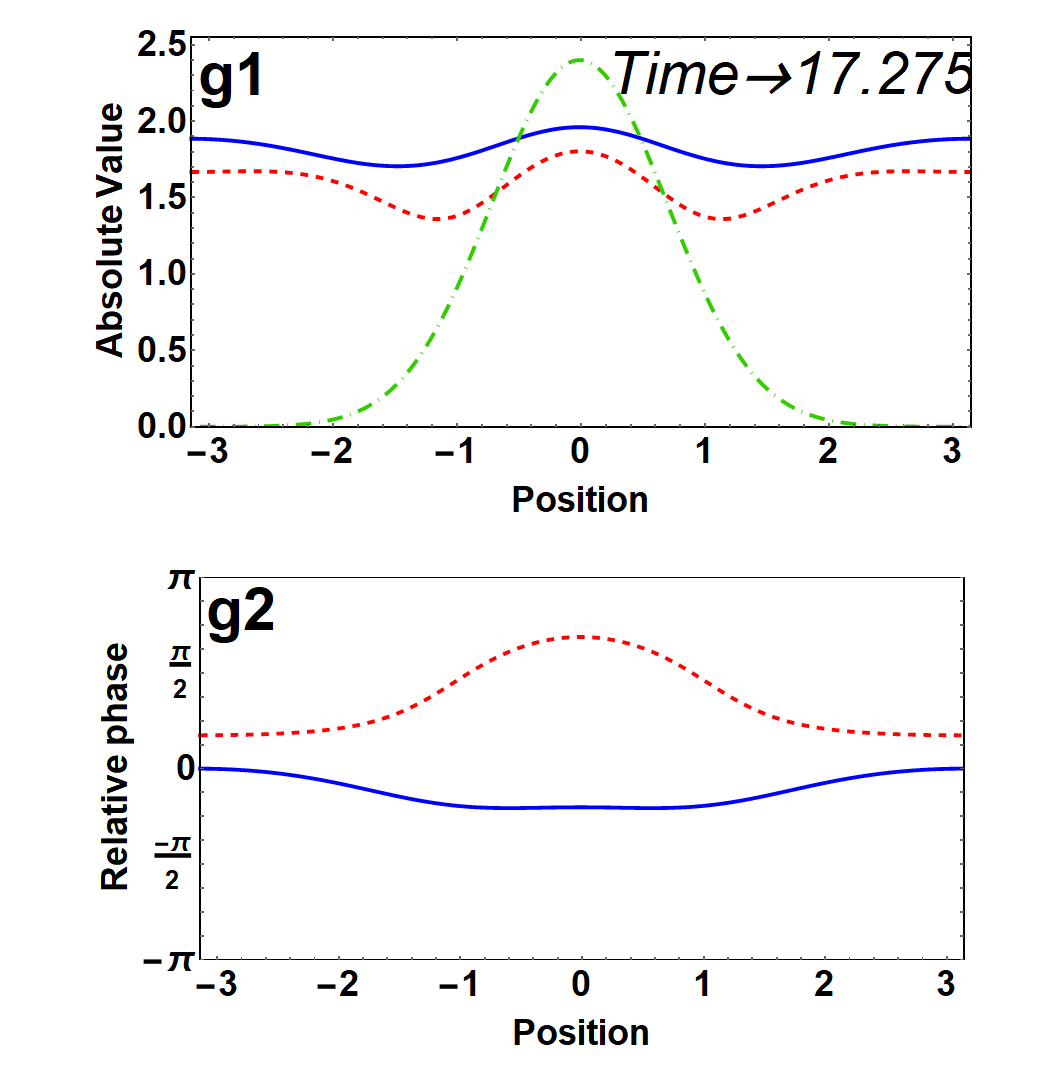}
    \includegraphics[scale=.11]{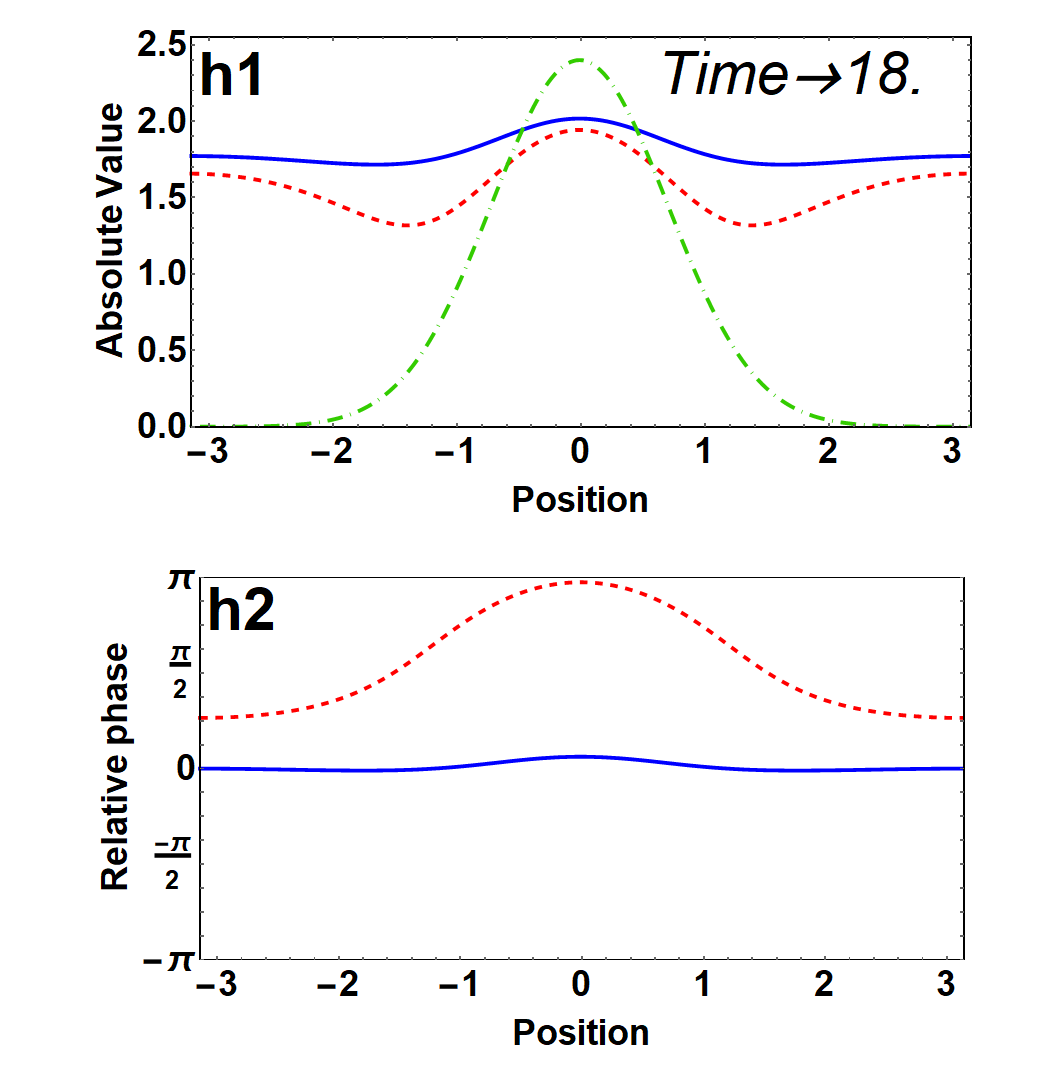}
    \caption{Snapshots of wavefunction propagation, representing half period of symmetric oscillations with coupling strength $J_0=4$, $w=1$, $\gamma=3$ and $\Gamma=1$.
     All frames are presented in pairs, where top frames (a1-h1) show absolute values of both wavefunctions (blue and red curves) and rescaled coupling potential (green curve), while bottom frames (a2-h2) show relative phases in both rings. Phases are plotted so that point $x=\pi$ for blue curve is fixed at 0, to eliminate phase oscillations term.}
    \label{fig:4}
\end{figure}

\begin{figure}
\centering
\includegraphics[scale=.11]{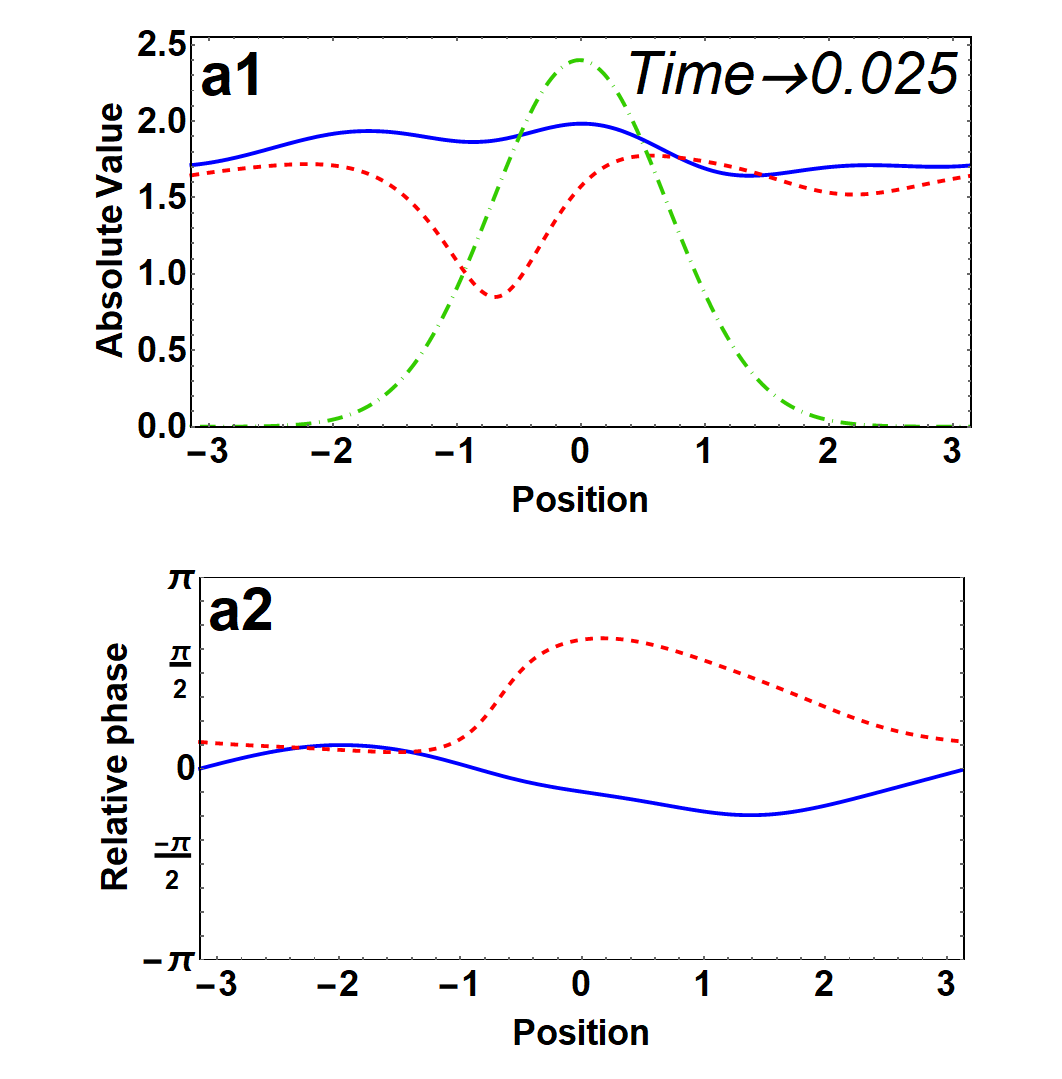}
\includegraphics[scale=.11]{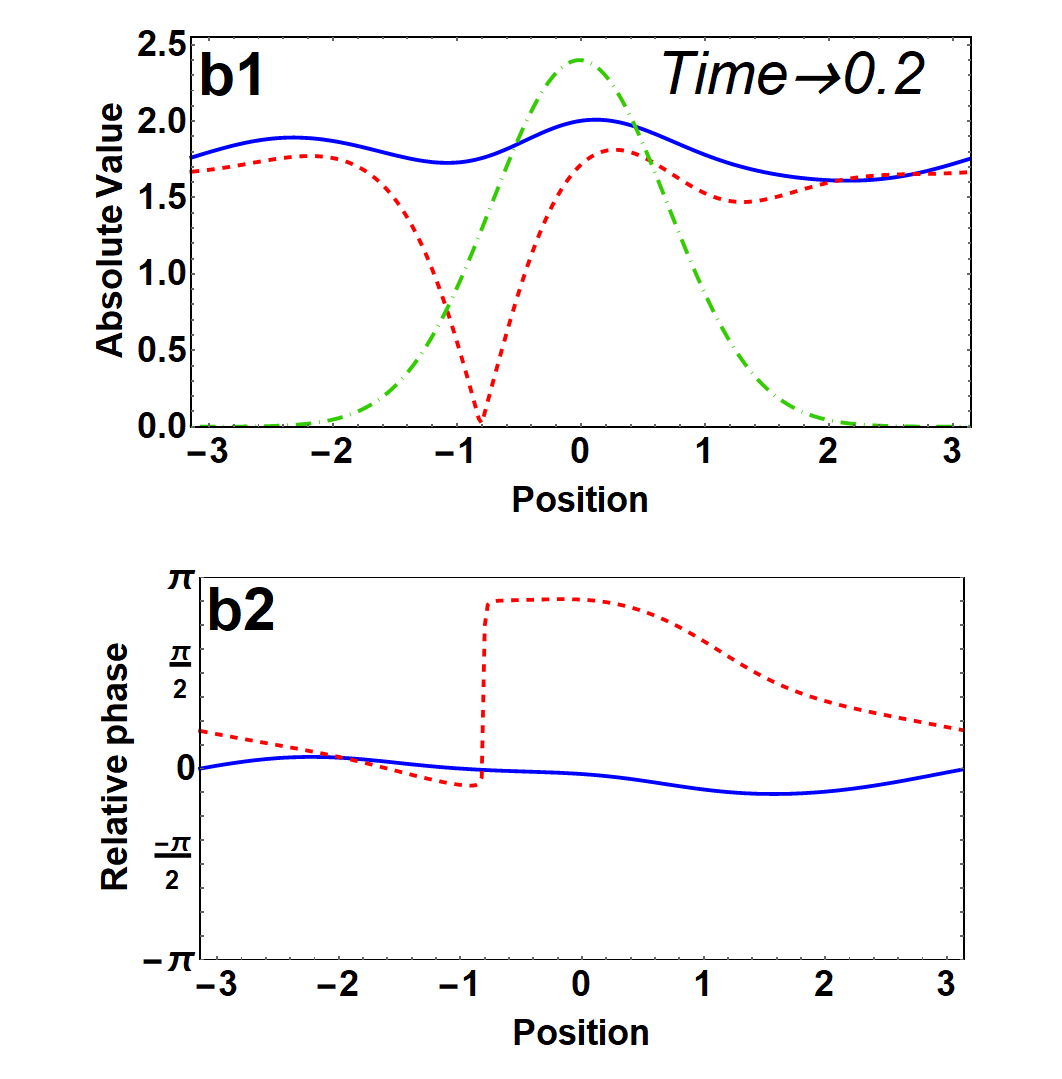}
\includegraphics[scale=.11]{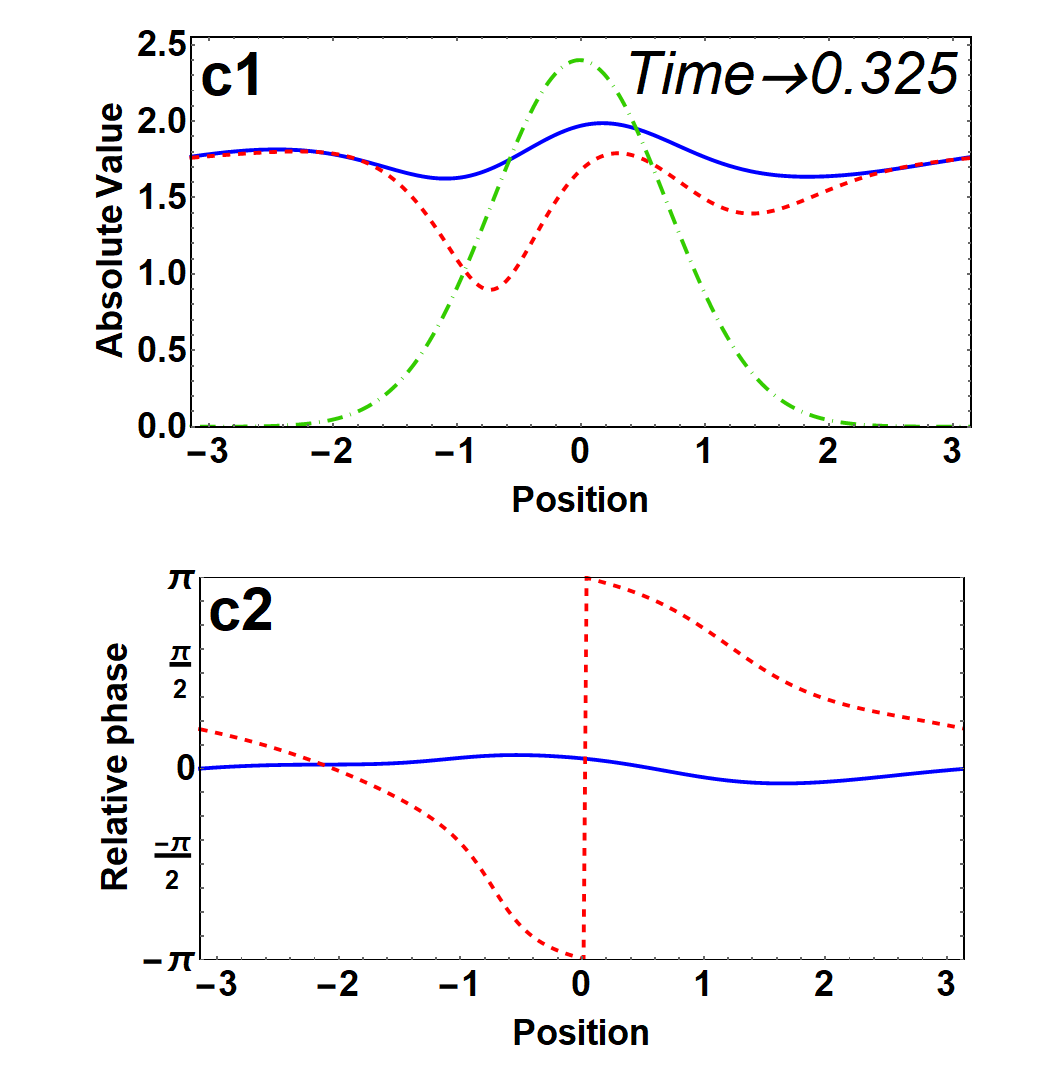}
\includegraphics[scale=.11]{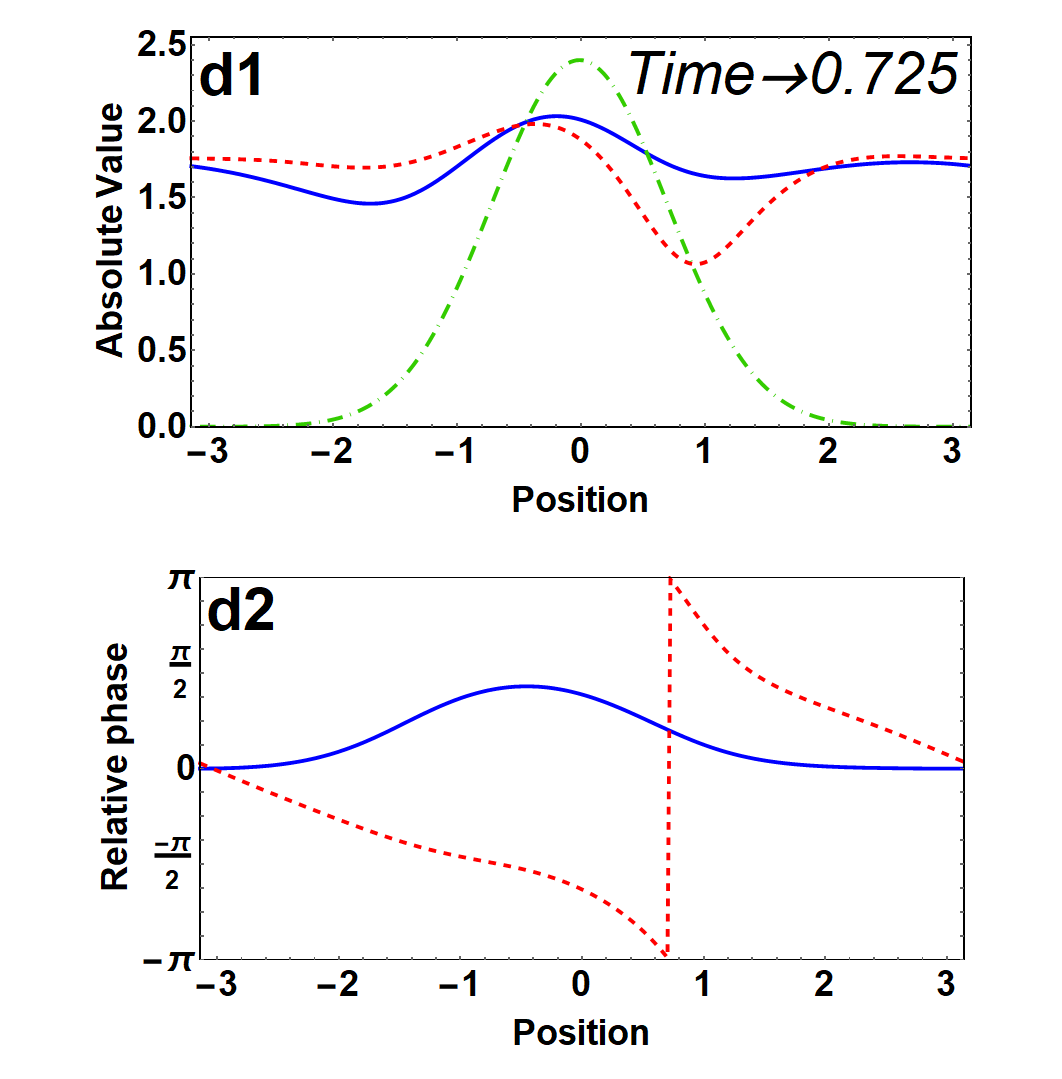}\\
\includegraphics[scale=.11]{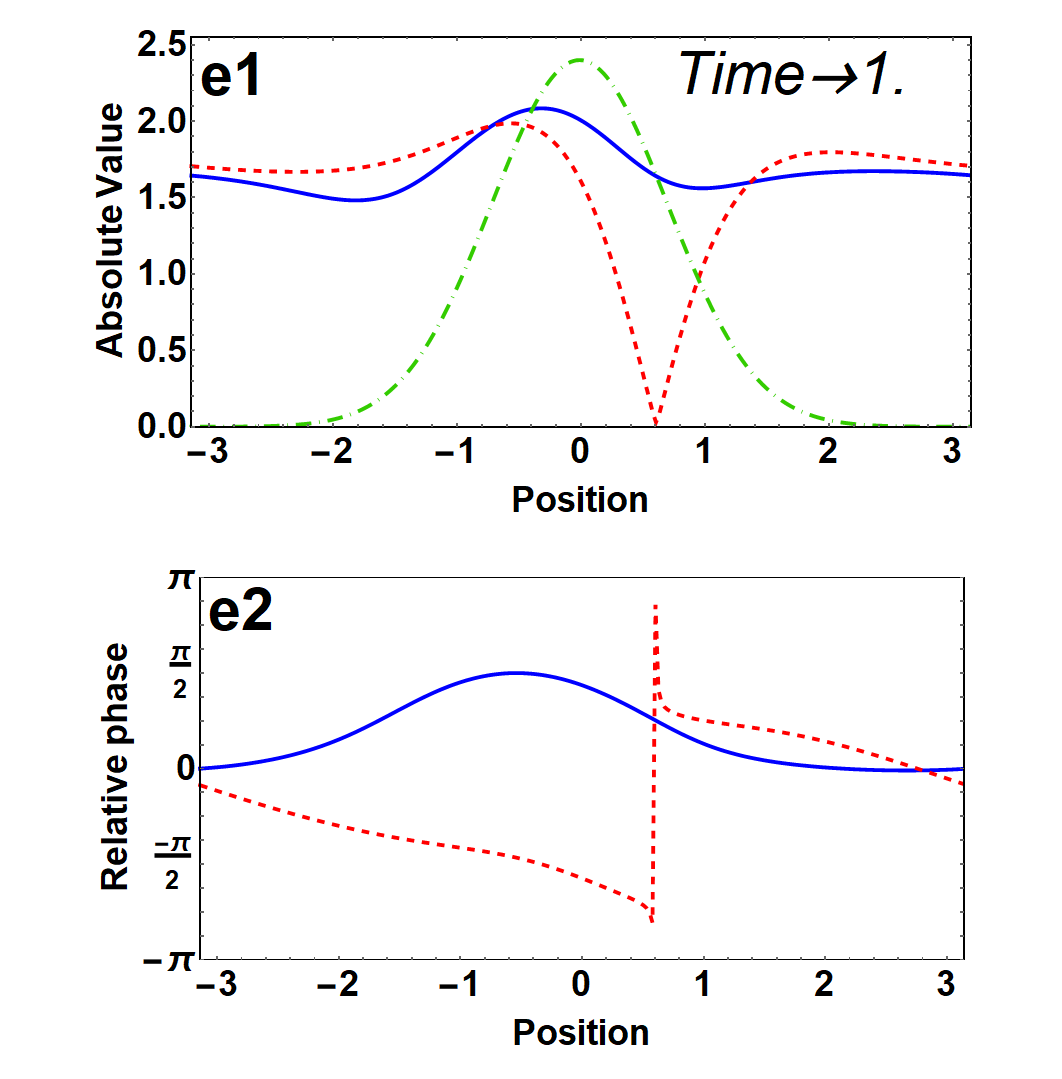}
\includegraphics[scale=.11]{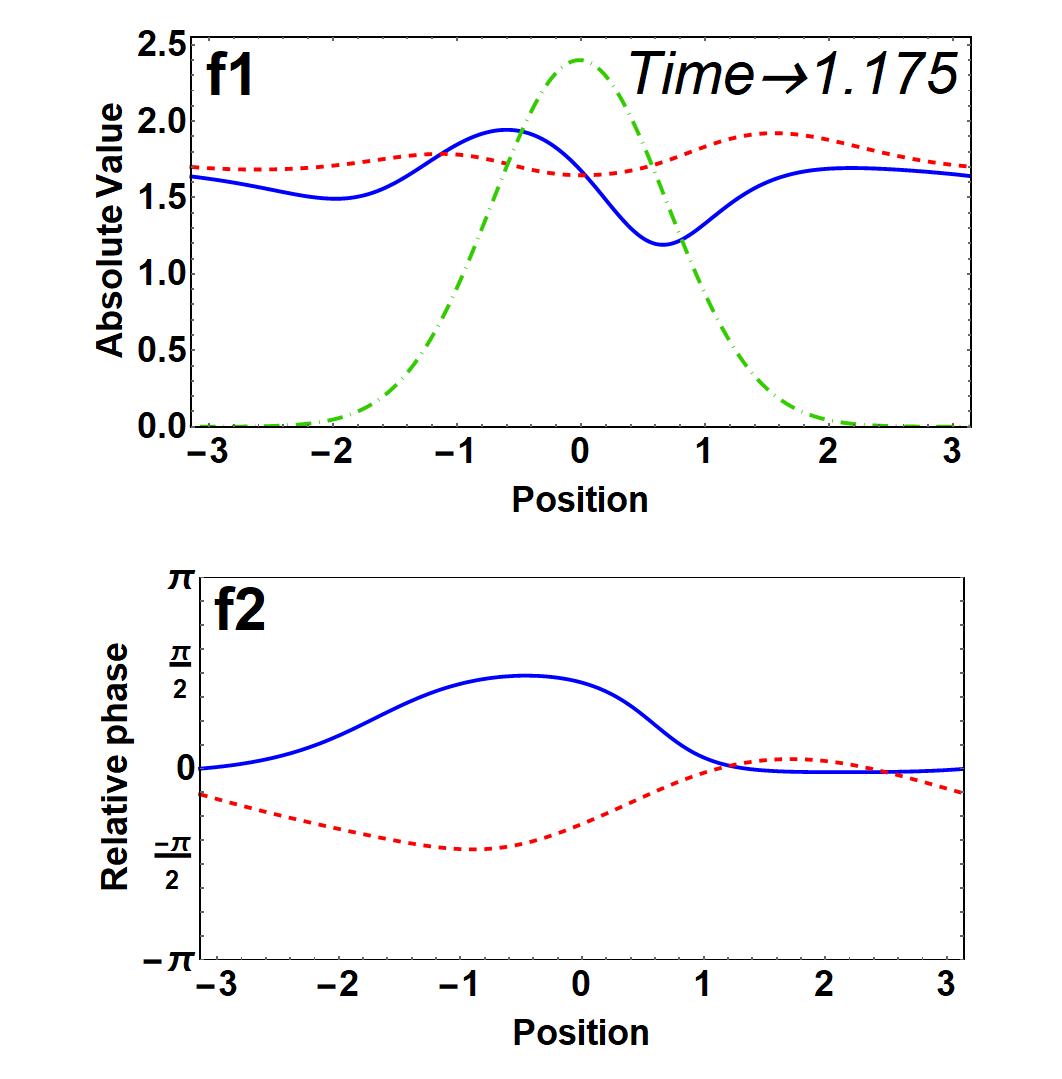}
\includegraphics[scale=.11]{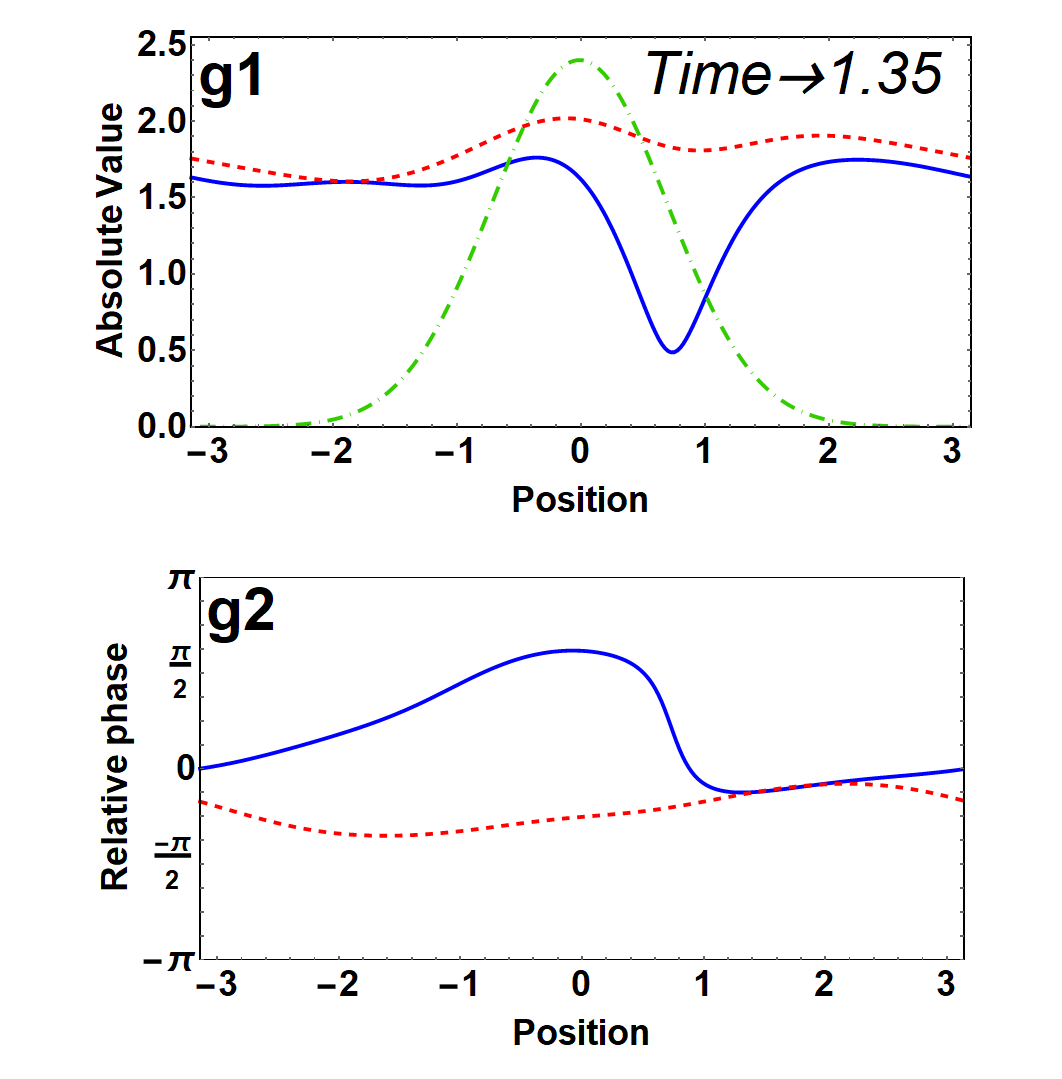}
\includegraphics[scale=.11]{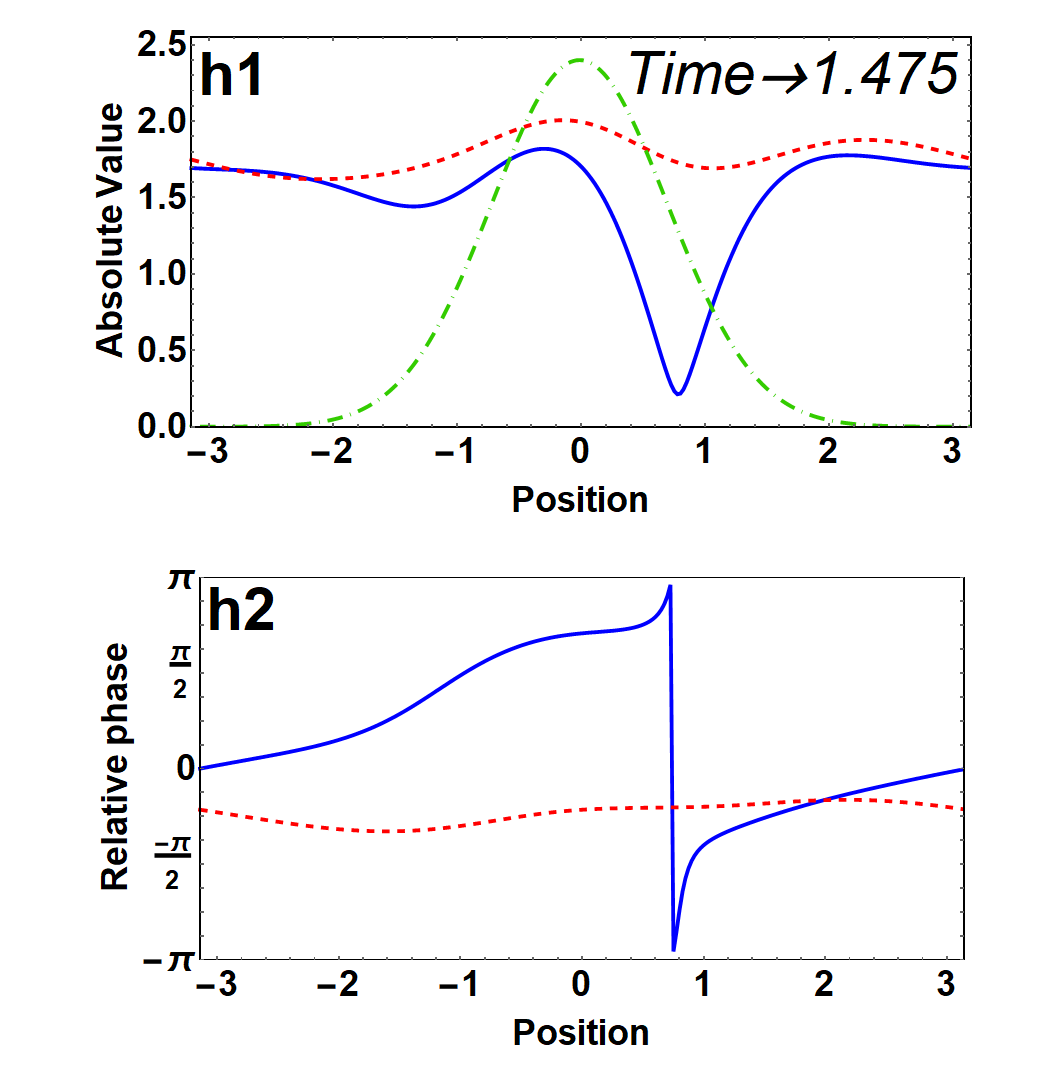}\\
\includegraphics[scale=.11]{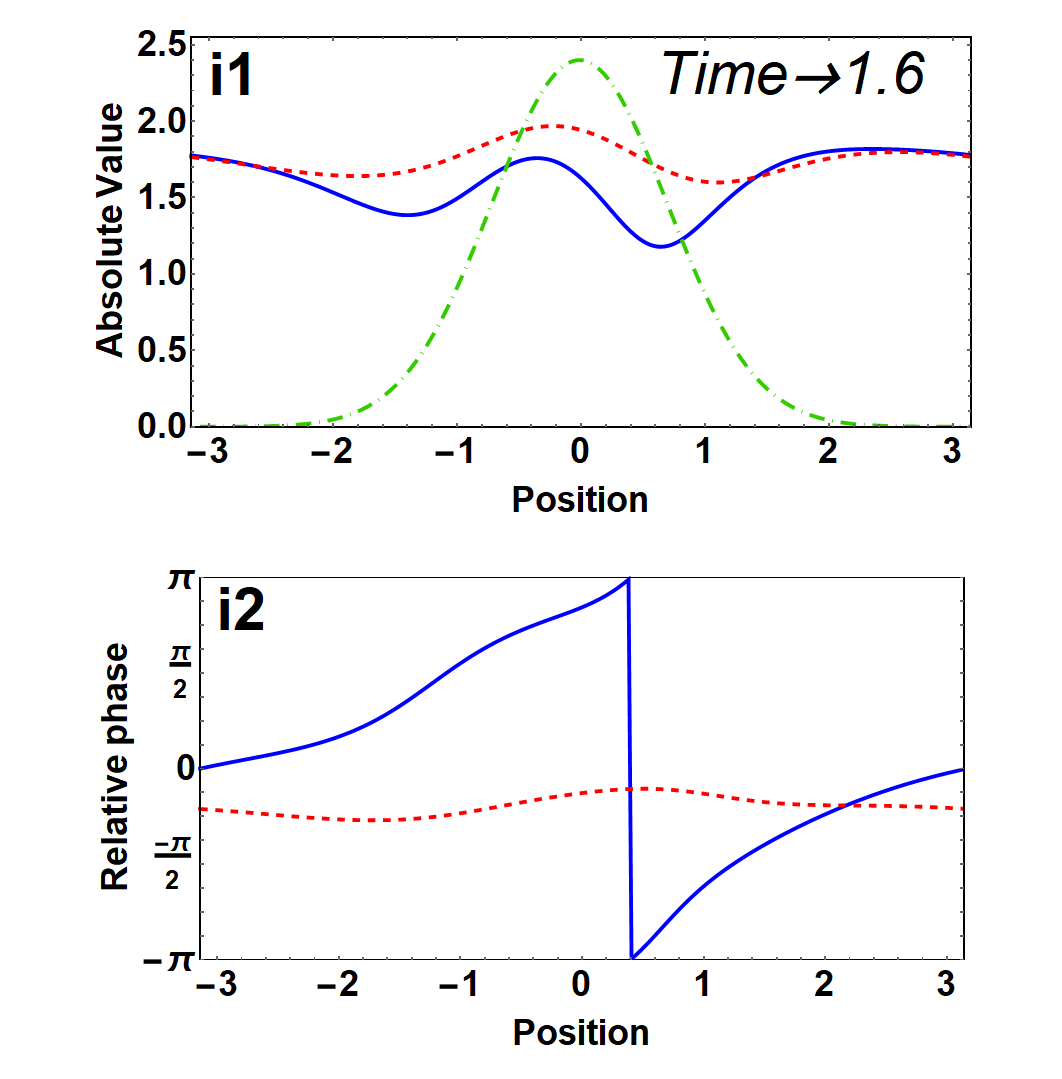}
\includegraphics[scale=.11]{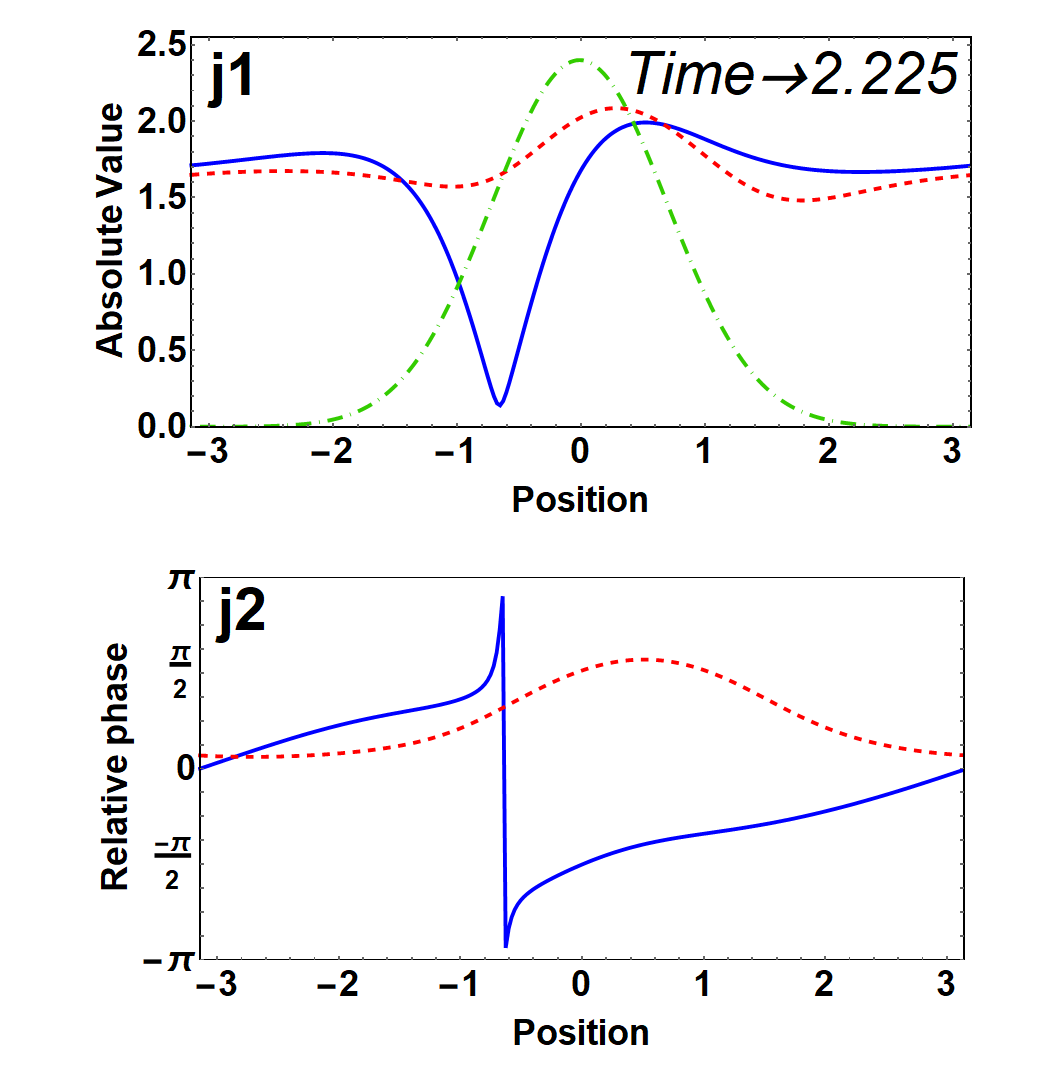}
\includegraphics[scale=.11]{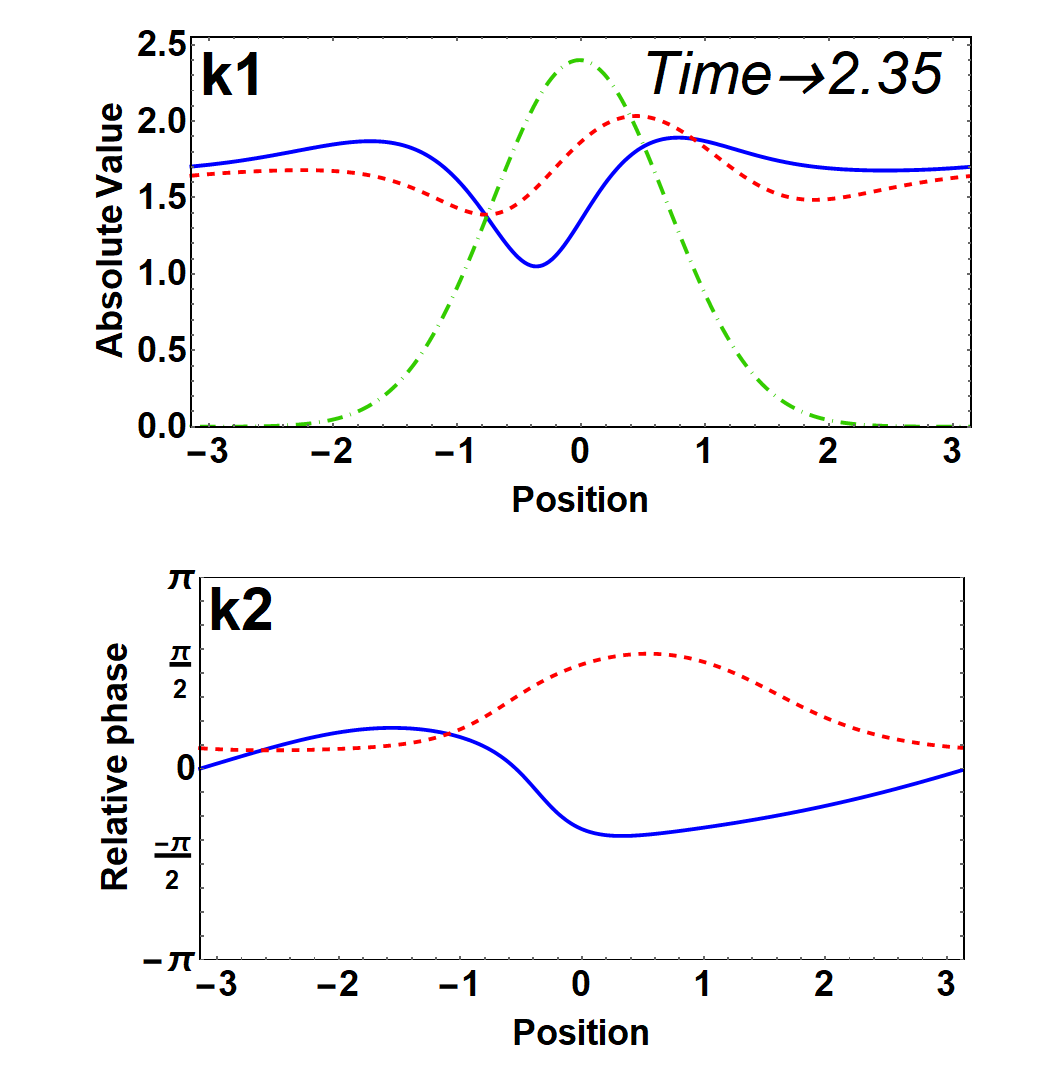}
\includegraphics[scale=.11]{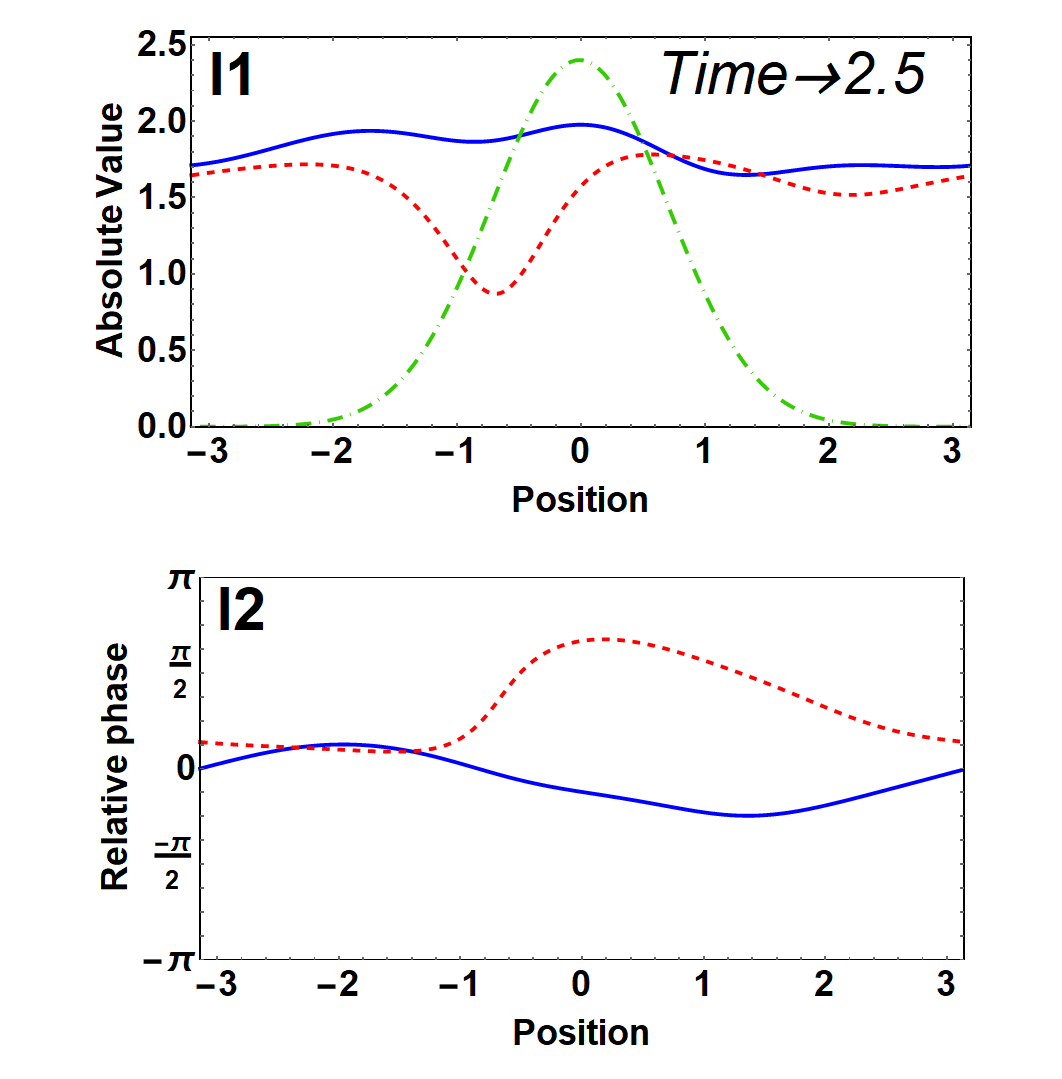}
\caption{Snapshots of wavefunction propagation, representing full period of asymmetric oscillations with coupling strength $J_0=5$ in the broad coupling regime, $w=1$, $\gamma=3$ and $\Gamma=1$. All frames are presented in pairs; top frames (a1-l1) show absolute values of both wavefunctions (blue and red curves) and rescaled coupling potential (green curve), bottom frames (a2-l2) show relative phases in both rings. Phase oscillations term was eliminated as in figure \ref{fig:4}.}   \label{fig:6}
\end{figure}

In the opposite limit, when the range of the coupling is comparable to the length of the ring (but not uniform yet) we also observe few classes of characteristic steady state dynamics and we classify them according to the (increasing) value of coupling strength $J_0$. We present results for $w=1$ and show asymptotic (in time) dynamics. In this case, since the most interesting steady state exhibits rather complex oscillations, corresponding to the limit cycle, we choose to present contour plots (see figure \ref{fig:3}) and dynamical snapshots at various times in figures \ref{fig:4} and \ref{fig:6}. Note that we always start from perturbed symmetric state given in Eq. (\ref{eq:4}), but our predictions are not sensitive to this particular choice. 

As it is expected, increase of the coupling strength leads to the increase of the oscillation frequency. An interesting observation, however, is that the transient period, i.e. time necessary for establishing oscillations, also increases with $J_0$. The oscillatory dynamics is symmetric with respect to both rings, with alternating field characteristics (amplitude and relative phase) exchanging after each half-period.

In the case of small coupling strength ($J_0 \lesssim 3.5$), the dynamics lead directly to the anti-symmetric state, analogous to the one presented in Figure \ref{fig:2}. This part is identical in both cases of broad and narrow coupling.

We did not find any asymmetric stationary states in the case of broad coupling. This class of states was only present for very narrow coupling, as mentioned above. Instead we observe directly transition to the next phase described below, namely symmetrically oscillating states. However, in the broad coupling case this is not the last category identified in our research. Above this class, as we describe below, there is an extra region of the most interesting solutions, where vortices are produced in alternating manner, in one or the other channel.

Upon increasing the value of coupling strength, at approximately $J_0 \simeq 3.8$ symmetric (with respect to the center of the Gaussian coupling function (\ref{eq:2})) oscillations begin to develop, initially with very long period, which becomes shorter at higher coupling strength. Details of the oscillations can be identified in figures \ref{fig:3} and \ref{fig:4}. In the first figure we now focus on the left panel (moving from the top to the bottom). First the contour plot of the time evolution of the amplitude and phase oscillations of $\psi_1$ is shown (wavefunction in the second channel is just shifted by half period), next we present the evolution of the norm in each channel ($N_i= \int |\psi_i|^2 dx$) as function of time and finally, we present more details, the evolution of the norm of each of the wavefunctions and the average within the full period of oscillations. 

These results are complemented by the full view of the wavefunction during its half-period oscillations in the asymptotic regime in figure \ref{fig:4}. The first and the third rows show the modulus of the wavefunctions (blue and red curves correspond to the first and second channels correspondingly), and second and fourth rows show the phase structure. We can trace the dynamics in which one of the channels develops two symmetric dips (see Fig.~\ref{fig:4} (a1) and (b1)), that develop slowly, reach bottom (c1) and eventually flatten to make exchange with its partner from the second channel (see frames (e1)-(h1)). In parallel we show the phase structure (frames (a2) though (h2)), and perhaps the most prominent feature is very steep phase profile in frames (c2)-(d2). It happens exactly at the time when the two dip structures in one of the moduli (blue curve in (c1)) reach zero at the minimum. Notice that the solution stays symmetric all the time. This will no longer be true when we go to the higher coupling regime.

\begin{figure}
\centering
\includegraphics[scale=.24]{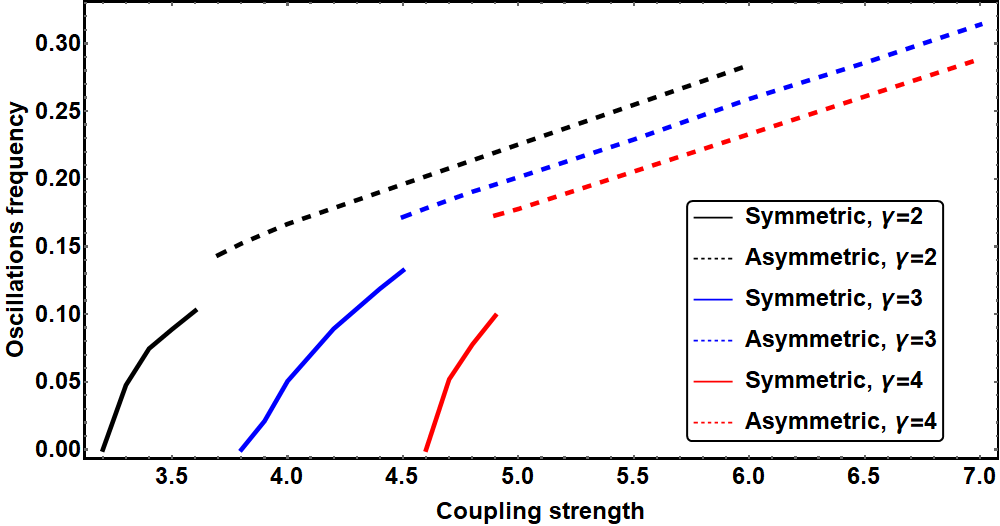}
\includegraphics[scale=.24]{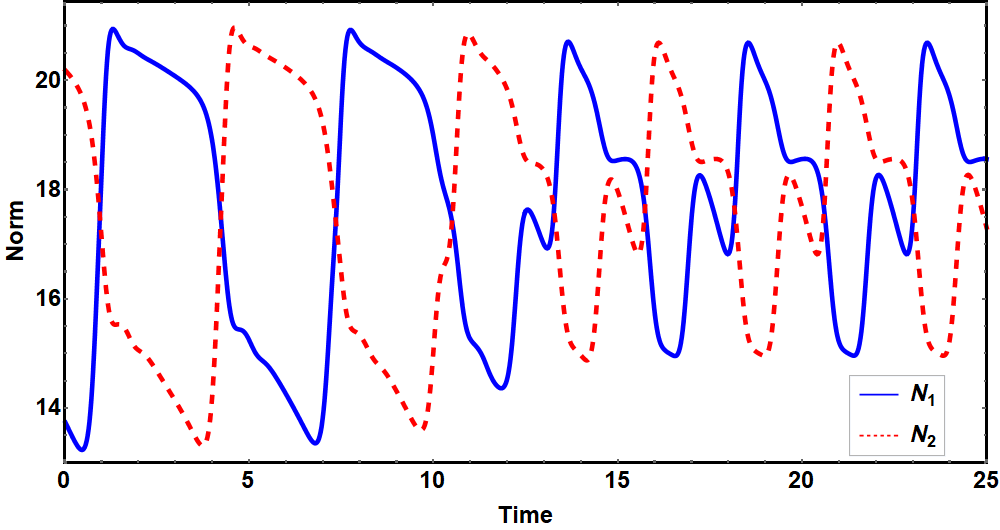}
\caption{Left panel: Frequency of the oscillations of the limit cycle solutions for three different values of $\gamma$. Symmetric oscillations, see Fig.~\ref{fig:4} are marked as continuous lines and asymmetric (see Fig.~\ref{fig:6}) with dashed lines. The other parameters: $w=1$, $\Gamma=1$. Right panel: The norm in both channels at $J_0 = 4.5$, where the blue line on left panel has discontinuity. We observe that after initial propagation (not shown) system develops into symmetrically oscillating state (shown in the time interval from $t=0$ to $t \approx 10$) and after several (typically 3 or 4, depending on initial perturbation) periods of oscillation system finally evolves into asymmetrical oscillations.
} \label{fig:5}
\end{figure}

 We investigated the frequency of oscillations of the periodic solutions. Results are presented in figure \ref{fig:5}. It is a collective plot containing the results not only for our central example of $\gamma = 3$, but also two different values of this parameter (notice that we keep the value of $\Gamma = 1$). At small value of the coupling frequency grows rapidly, then a sudden jump occurs and further growth is linear. This jump is associated with yet another bifurcation (symmetry breaking), and as a matter of fact solutions marked with dashed lines no longer belong to the class described above, as it exhibits asymmetric oscillations. We will describe it in more details in the next paragraph. In our leading example of $\gamma = 3$ this phase transition occurs at coupling $J_0 \approx 4.5$. In the transition region around this value of the coupling we observe that dynamics leads first to the symmetric oscillations and after the transient period of several symmetric oscillations it follows to the asymmetric oscillations, which are asymptotically stable. This transient behaviour can be identified in the right panel of figure \ref{fig:5}, where we plot norm of the wavefunctions.

For values of the coupling above $J_0 \approx 4.5$ (this particular value
corresponds to $\Gamma = 1$ and $\gamma = 3$, see fig.\ref{fig:5}) phase structure becomes asymmetric with respect to the center of the coupling function, see Eq.~(\ref{eq:2}). This leads to periodic, limited in time, appearance of vortex in one of the channels,
which then (on the regular basis) reappears in the opposite channel,
with inverted topological charge. The whole dynamics has again a
form of regular oscillations and we show various phases of the
evolution of the wavefunctions in figures \ref{fig:3} and
\ref{fig:6}. In Fig.~\ref{fig:3} we show the time evolution of the
modulus of the wavefunction and its phase on the contour plot. Note
that in this regime there is single dip that appears once when a
vortex is created (vortex in our case, as we mentioned above is
equivalent to excitation) and again when vortex disappears, only to
show up a bit later in the opposite ring. One can follow this
process even closer looking at figure \ref{fig:6}. In panel (a1) we
see the dip which is just about to reach zero at around $x=-1$. The
phase around this point is very steep (panel (b)), and vortex is
created (there is a phase across the ring equal to $2 \pi$). Then
the wavefunction flattens, until second dip starts to develop at
around $x=1$. Once second dip reaches zero, vortex disappears. Then
the whole process repeats itself in the second ring in reversed
order (panels (g)-(l)). This type of behavior was not observed in
the case of narrow coupling, when we only have a symmetric
structure. It seems that there is some distance between the edges of
the coupling function necessary for this spacial symmetry to be
broken.

\section{On vortex creation without stirring.}
\label{sec:vortex}

Our system is non-Hermitian and as such it does not conserve
topological charge; vortices can be created during the dynamics,
even if the system is rotationally invariant. It happens due to the
modulational instability, and as we mentioned above in this 1D
system vortex is equivalent to higher momentum excitations.
Previously \cite{SciRep}, for some values of the coupling constant
we demonstrated that the system can, starting from perturbed
symmetric state, arrive at the stationary states defined as

\begin{eqnarray}
\label{vortex1}
\Psi_1(x,t)= - \Psi_2(x,t)=\sqrt{\frac{\gamma}{\Gamma}}
e^{i\left[\kappa x-(\frac{\gamma}{\Gamma}+\kappa^2-c)t\right]}.
\end{eqnarray}
Here $\kappa$ is an integer number expressing the value of the
topological charge. This antisymmetric state is stable against
initial perturbations. There is also symmetric solution, where
$\Psi_1(x,t)=\Psi_2(x,t)$, but they are usually unstable and we
never observed that in the asymptotic limit of our simulations.

\begin{figure}
\centering
\includegraphics[scale=.11]{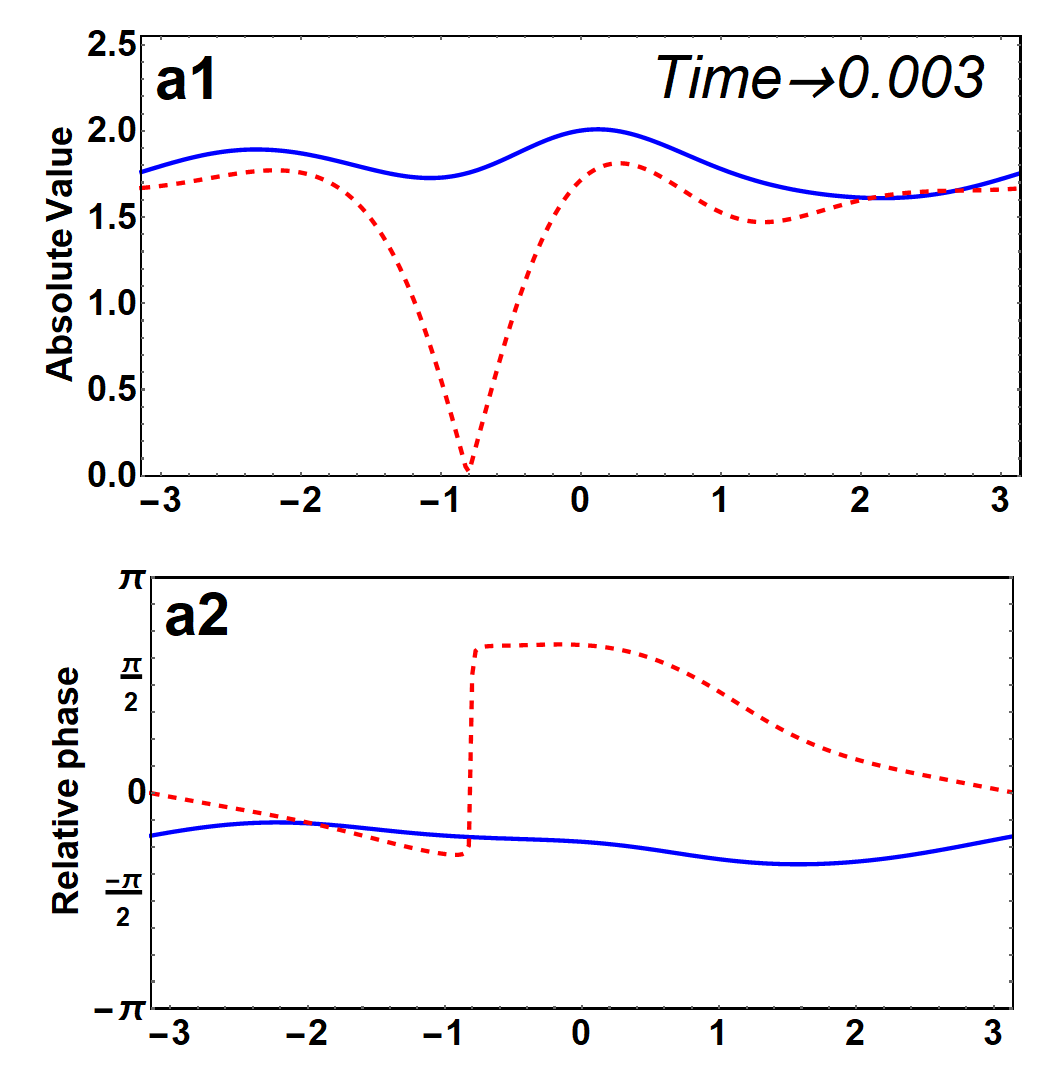}
\includegraphics[scale=.11]{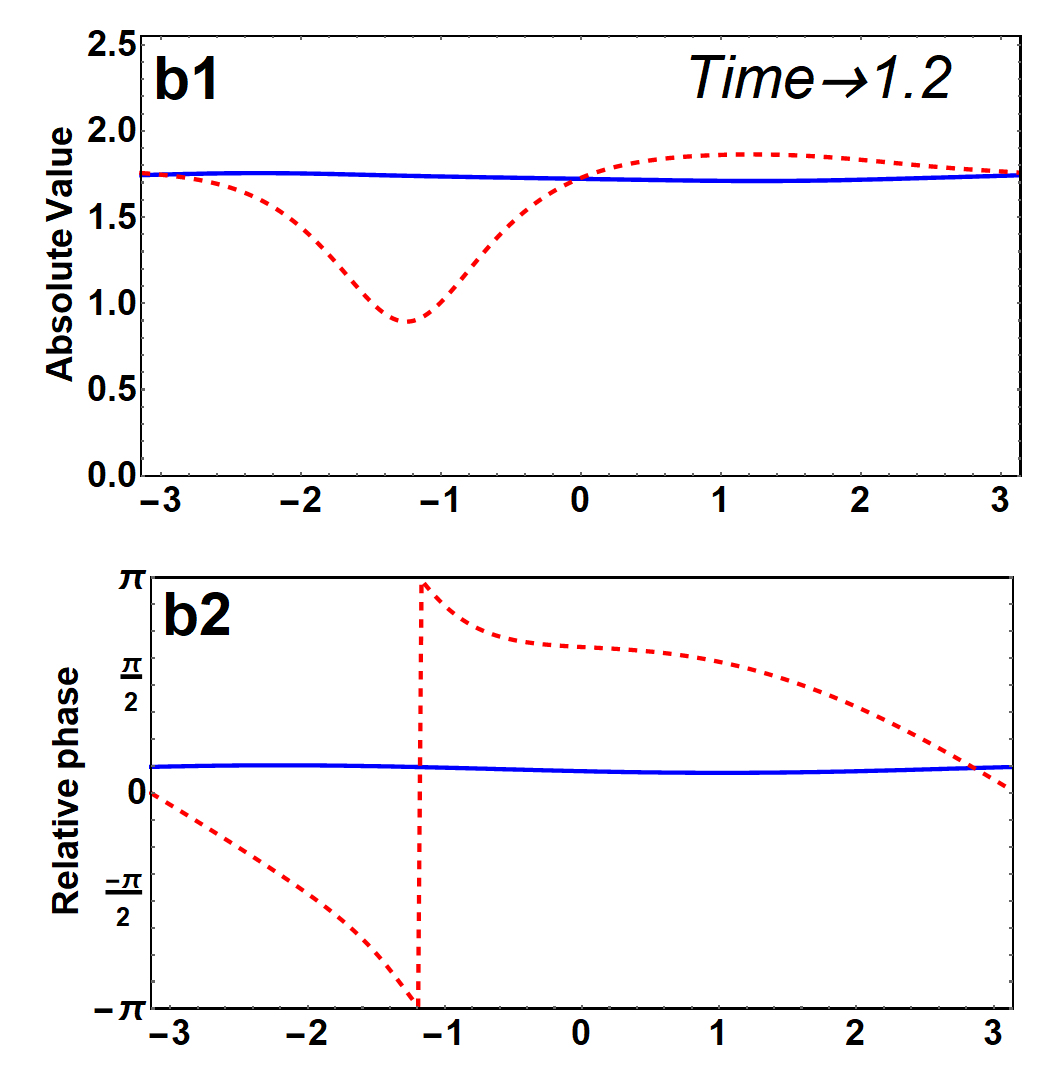}
\includegraphics[scale=.11]{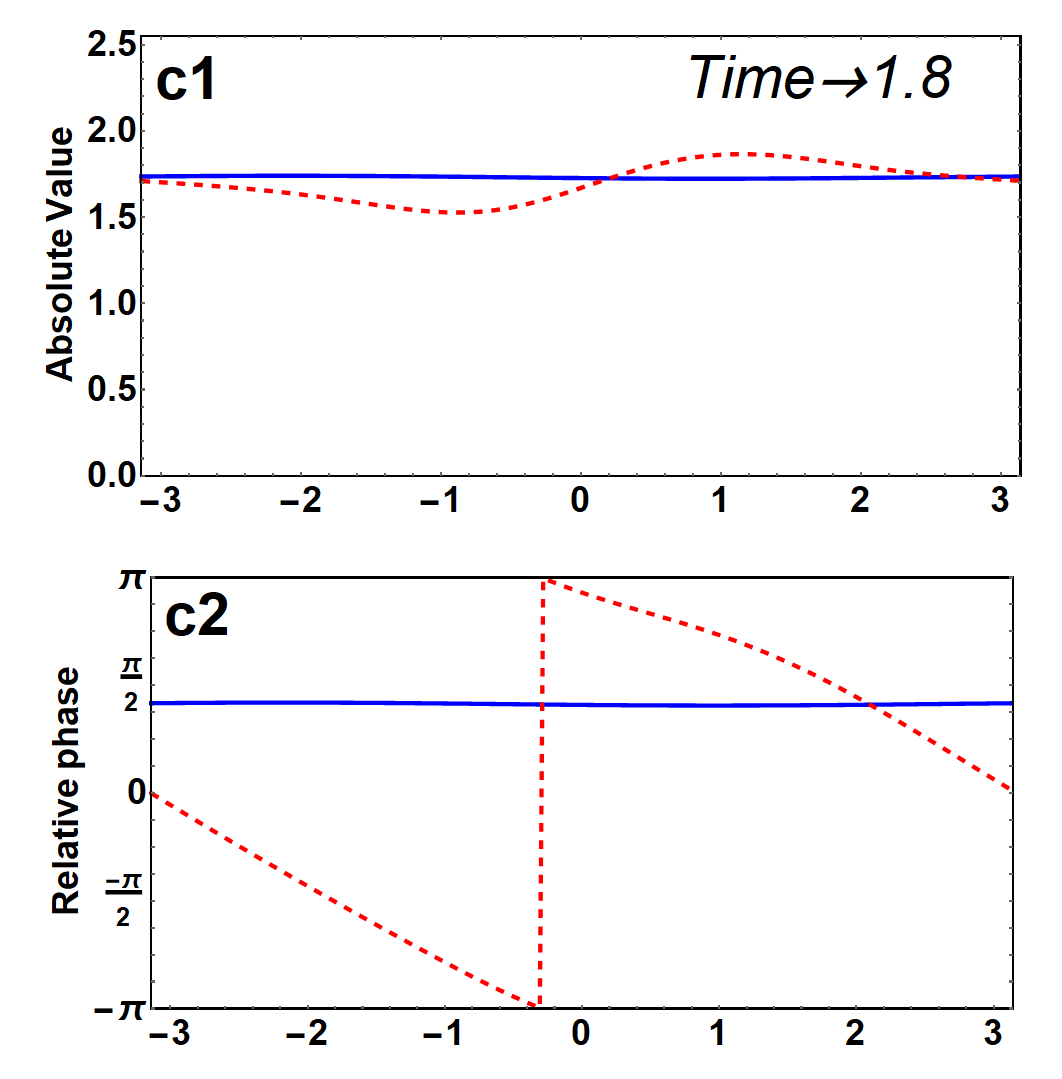}
\includegraphics[scale=.11]{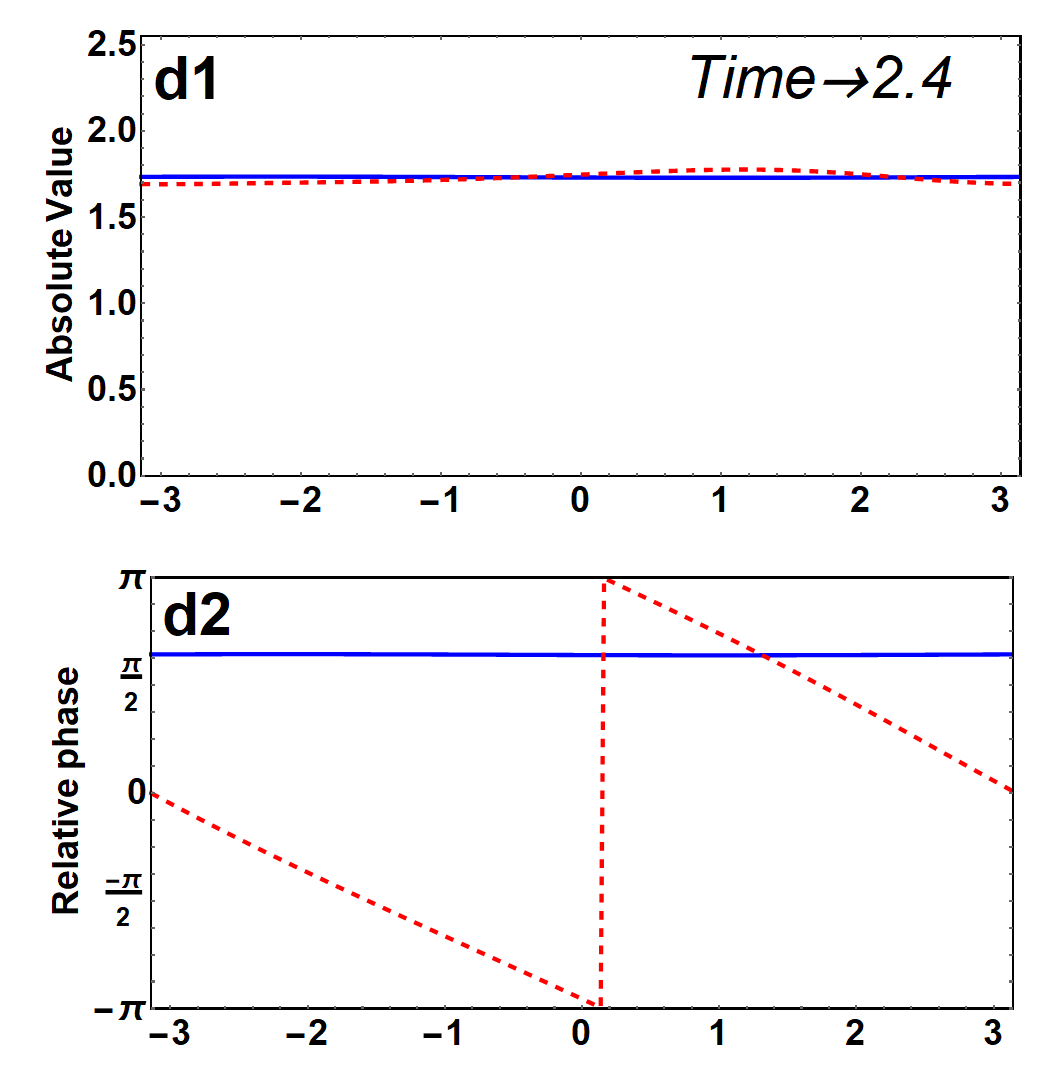}\\
\caption{Snapshots of wavefunction propagation, representing relaxation from point of vortex generation (point \emph{b} from Figure \ref{fig:6}) after turning off the coupling, for $\gamma=3$ and $\Gamma=1$. All frames are presented in pairs; top frames (a1-d1) show absolute values of both wavefunctions (blue and red curves), bottom frames (a2-d2) show relative phases in both rings. Phase oscillations term was eliminated as in figure \ref{fig:4}.} \label{fig:7}
\end{figure}

So far we could create vortices with equal topological charges in both channels. But here, when we define inhomogeneous local coupling, additionally varying in time (it is enough to switch it on for some time and then switch it off) there is even more exciting possibility of creating vortex only in one ring, accompanied by constant solution in the other. The idea is based on the results obtained for the broad coupling function, described in Sec.~\ref{sec:broad}. Imagine that we start from perturbed symmetric state (as we did in all the cases described in this manuscript) and we allow the system to evolve towards the regime of asymmetric oscillations. When the system is already in this regime, we will abruptly turn off the coupling between rings. Dynamics corresponding to this scenario is illustrated in figure \ref{fig:7}. Here we present series of snapshots of both wavefunctions and their phases. In Fig.~\ref{fig:6} in panels (a1) and (b1) the coupling is still on, and one of the wavefunctions is at the stage when vortex is created and dip in the amplitude is fully developed (reaches the bottom). In the series of panels (columns) in figure \ref{fig:7} the coupling between rings is off and we observe slow, independent relaxation in each of them separately. In this circumstances one of the wavefunctions preserves its vortex structure and tends to the solution with smooth modulus, defined in Eq.~(\ref{vortex1}) with $J_0=0$, and the opposite ring develops wavefunction defined by Eq.~(\ref{eq:3}). Turning off coupling in any other moment, while non-zero topological charge is present in the system, leads to even faster relaxation into described state. In this scenario we showed how to selectively create vortex in one ring. In the remaining part of the manuscript we present experimental proposal, where such dynamics may be investigated, and propose similar coupled ring systems that we intend to investigate in the near future.

\section{Experimental proposal: plasmonic and metamaterial effects in arrays of nanorings}
\label{sec:experiment}

\begin{figure}
\centering
\includegraphics[scale=.45]{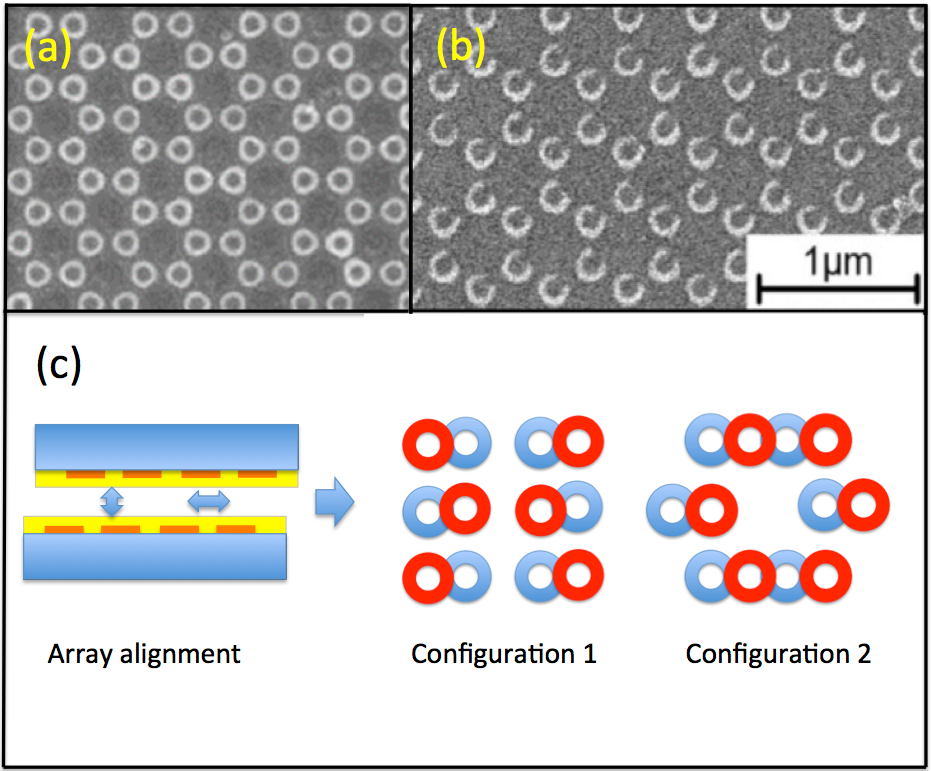}
\caption{(a) and (b) SEM images of the circular and c-shaped nanorings produced by SNL. (c) Schematic of the assembly of the coupled nanoring arrays (left panel), and the resulting examples of possible offset-dependent ring alignment configurations. } \label{fig:8}
\end{figure}

The effects discussed above can be studied via electromagnetic response of arrays of metallic nanorings. These form photonic/plasmonic crystals with enhanced electromagnetic response \cite{wang}. Such structures can be made by a variety of techniques, including the electron beam lithography (EBL), as well as the Shadow Nanosphere Lithography (SNL) \cite{giersig1,giersig2}. This last technology provides an inexpensive route to complex periodic nanostructures, including nanorings. Figures \ref{fig:8} (a) and (b) show scanning electron microscopy (SEM) images of arrays of metallic nanorings (circular and c-shaped, respectively) deposited on a substrate using SNL. Such arrays can be used as a basis for corresponding arrays of coupled nanorings, which will have a very pronounced electromagnetic response depending on the physics of the inter-ring coupling and non-linear, intra-ring losses. To obtain the coupled nanoring arrays, we make two copies of an array (e.g. circular rings), as shown schematically in figure \ref{fig:8} (c), left panel. Each copy is an array, deposited on a transparent but lossy substrate (blue), and with the rings (orange) coated with a dielectric film (yellow). The two copies are sandwiched together, but depending on their relative offset, various configuration overlaps can be achieved. For example, one can perfectly align the rings (not shown) and therefore realize a uniform coupling ($J(x) = constant$). However, by shifting the two arrays relative to each other, one can achieve localized coupling as discussed above (configuration 1), or multiple localized coupling (configuration 2). In configuration 2 one has a more complicated situation, with some rings coupled only to single rings in the other array, and some to multiple rings (in both arrays). In addition, even in the configuration 1, one can achieve a pair of localized coupling regions by making horizontal shift adjustments. The local coupling can be also realized by tilting rings with respect to each other, since the strength of the coupling is proportional to the distance. One can also use inhomogeneous filling of the inter-ring space.

Experimentally realizable, more complex, configurations could be of interest, and we will study the corresponding models elsewhere.

The physics described in previous sections relies on sufficiently large non-linear absorption losses. These can be controlled by choosing highly non-linear materials for the substrates (or substrate coatings), such as organic semiconductors, e.g. acetoacetanilide \cite{babu}. Another way to control the intra-ring nonlinear absorption would be to employ lossy metals for the body of the ring, or use the c-shaped rings, as shown in figure \ref{fig:8} (b), with a nonlinear material coating in the opening. Such processing is possible \cite{giersig3}.
The inter-ring coupling can be easily controlled via the dielectric coating (yellow colored layer in the left panel of figure \ref{fig:8} (c)). Also, if the substrates could be made conducting, the bias across the pair could control the inter-ring coupling as well, relying on the non-linearity of the current-voltage characteristics of these metal-insulator-metal (MIM) structures.

\section{Conclusions}

In this work we continued our investigation of simple ring-shaped nonlinear waveguides in the presence of the linear gain and nonlinear dissipation, adding local coupling between two channels. We have identified stationary and dynamic solutions both in case of narrow and broad coupling. These solutions represent different types of symmetry breaking, corresponding to bifurcations of fixed points (stationary solutions) and of limiting cycles (symmetric and asymmetric oscillatory solutions).  Dynamic solutions, connected with vortex generation, allow us both to control channel populations and generate vortex states in the system. We propose experimental realization of our model in nanoplasmonic system.

\acknowledgments{The work was supported by the Polish National
Science Centre 2016/22/M/ST2/00261 (A. Ramaniuk and M. Trippenbach.)
N.V.H. was supported by Vietnam National Foundation for Science and Technology
Development (NAFOSTED) under grant number 103.01-2017.55. M.G. acknowledges the funding by the Guangdong Innovative and Entrepreneurial Team Program titled "Plasmonic Nanomaterials and Quantum Dots for Light Management in Optoelectronic Devices" (No.2016ZT06C517).\\
The authors declare no conflict of interest.}



\renewcommand\bibname{References}

\end{document}